\documentclass[twocolumn]{aastex631}

\usepackage{comment}
\usepackage{graphicx}
\usepackage{txfonts}

\newcommand{\rev}[1]{{\textcolor{black}{#1}}}
\newcommand{\rerev}[1]{{\textcolor{black}{#1}}}
\newcommand{\eqref}[1]{(\ref{#1})}

\defcitealias{Ohno&Fortney22a}{Paper I}

\shorttitle{Tracing Nitrogen In Giant Planet Atmospheres II}
\graphicspath{{./}{figures/}}
\begin{document} 

\title{Nitrogen as a Tracer of Giant Planet Formation. II.: Comprehensive Study of Nitrogen Photochemistry and Implications for Observing NH$_3$ and HCN in Transmission and Emission Spectra}

\shortauthors{Ohno \& Fortney}

\author[0000-0003-3290-6758]{Kazumasa Ohno}
\affiliation{Division of Science, National Astronomical Observatory of Japan, 2-21-1 Osawa, Mitaka-shi, Tokyo, Japan}
\affiliation{Department of Astronomy \& Astrophysics, University of California, Santa Cruz, 1156 High St, Santa Cruz, CA 95064, USA}

\author[0000-0002-9843-4354]{Jonathan J. Fortney}
\affiliation{Department of Astronomy \& Astrophysics, University of California, Santa Cruz, 1156 High St, Santa Cruz, CA 95064, USA}

\begin{abstract}
\rev{Atmospheric nitrogen may provide important constraints on giant planet formation.}
\rev{Following our semi-analytical work \citep{Ohno&Fortney22a}, we further pursue the relation between observable NH$_3$ and an atmosphere's bulk nitrogen abundance by applying the photochemical kinetics model VULCAN across planetary equilibrium temperature, mass, age, eddy diffusion coefficient, atmospheric composition, and stellar spectral type. We confirm that the quenched NH$_3$ abundance coincides with the bulk nitrogen abundance only at sub-Jupiter mass ($\la1M_{\rm J}$) planets and old ages ($\ga1~{\rm Gyr}$) for solar composition atmospheres, highlighting important caveats for inferring atmospheric nitrogen abundances. Our semi-analytical model reproduces the quenched NH$_3$ abundance computed by VULCAN and thus helps to infer the bulk nitrogen abundance from a retrieved NH$_3$ abundance. 
}
\rev{By computing transmission and emission spectra, we predict that the equilibrium temperature range of $400$--$1000~{\rm K}$ is optimal for detecting NH$_3$ because NH$_3$ depletion by thermochemistry and photochemistry is significant at hotter planets whereas entire spectral features become weak at colder planets.
For Jupiter-mass planets around Sun-like stars in this temperature range, NH$_3$ leaves observable signatures of $\sim50~{\rm ppm}$ at $1.5$, $2.1$, and $11~{\rm {\mu}m}$ in transmission spectra and $>300$--$100$ ppm at $6~{\rm {\mu}m}$ and $11~{\rm {\mu}m}$ in emission spectra. The photodissociation of NH$_3$ leads HCN to replace NH$_3$ at low pressures.  However, the low HCN column densities lead to much weaker absorption features than for NH$_3$.  The NH$_3$ features are readily accessible to JWST observations to constrain atmospheric nitrogen abundances, which may open a new avenue to understand the formation processes of giant exoplanets.}

\end{abstract}
\section{Introduction}
The composition of giant planet atmospheres offer valuable clues to the planet formation process.
Over the past decade a number of studies have suggested that atmospheric elemental ratios, such as the carbon-to-oxygen ratio (C/O), can diagnose a giant planet's formation location and accretion history within a disk \citep[e.g.,][]{Oberg+11,Oberg&Bergin16,Oberg&Wordsworth19,Madhusudhan+14,Madhusudhan+17,Ali-dib+14,Helling+14,Thiabaud+15,Piso+15,Piso+16,Espinoza+17,Eistrup+16,Eistrup+18,Eistrup+22,Cridland+16,Cridland+17,Cridland+19,Booth+17,Booth&Ilee19,Ohno&Ueda21,Turrini+22,Schneider&Bitsch21,Molliere+22,Pacetti+22,Bitsch+22,Eistrup22}.
\rev{Recent studies also suggest that nitrogen provides valuable insights on the planet formation processes \citep{Piso+16,Cridland+20,Ohno&Ueda21,Turrini+22,Notsu+22}. Several studies have discussed the formation environment of Jupiter in our Solar System based on atmospheric nitrogen and noble gas abundances \citep[e.g.,][]{Owen+99,Gautier+01,Guillot&Hueso06,Monga&Desch15,Ali-dib17,Mousis+19,Oberg&Wordsworth19,Bosman+19,Ohno&Ueda21,Aguichine+22}.}

\rerev{NH$_3$ and HCN are likely the most easily accessible nitrogen species in giant exoplanet atmospheres via spectroscopy \citep{MacDonald&Madhusudhan17}, as the remaining main nitrogen reservoir N$_2$ has negligibly low visible and infrared opacity}. 
However, constraining an atmosphere's bulk nitrogen abundance from NH$_3$ and HCN is a complex task.
It is well known from previous work that the NH$_3$ and HCN abundances in the observable atmosphere readily deviates from thermochemical equilibrium abundances due to vertical mixing and photochemistry \citep[e.g.,][]{Moses+11,Line+11,Venot+13}. 
For warm planets with $T_{\rm eq}\la1200~{\rm K}$, \citet{Fortney+20} investigated disequilibrium NH$_3$ abundances on Saturn-like planets across $T_{\rm eq}$-space and found that the NH$_3$ abundance depends on a number of factors, such as planetary mass, age, and metallicity.
\rev{ \citet{Ohno&Fortney22a} (hereafter \citetalias{Ohno&Fortney22a}) further generalized the conditions under which the vertically quenched NH$_3$ abundance coincides with the bulk nitrogen abundance, based on a suite of radiative-convective atmosphere models and semi-analytical arguments.
\citetalias{Ohno&Fortney22a} suggested that the NH$_3$ abundance coincides with the bulk nitrogen abundance only at sub-Jupiter planetary mass ($\la 1M_{\rm jup}$), old age ($\ga 1$ Gyr), and low atmospheric metallicity ($< 10\times$ solar value); otherwise, N$_2$ dominates over NH$_3$, making the observable NH$_3$ abundance only a lower limit of bulk nitrogen abundance.
}

\rev{\citet{Fortney+20} and \citetalias{Ohno&Fortney22a} relied on the so-called quench approximation that estimates the vertically quenched NH$_3$ abundance from timescale arguments; however, the quench approximation is not always accurate. 
While the quench approximation yields vertically-constant abundance of disequilibrium species, \citet{Tsai+18} showed that the abundance can vary with altitudes even above the quench pressure level, especially for hot atmospheres. 
\citet{Molaverdikhani+19} also pointed that molecular diffusion and photodissociation can cause vertically nonuniform abundance profiles.
}
In fact, \citet{Hu21} investigated photochemistry on template/cold H$_2$-rich planets and found that NH$_3$ tends to be readily depleted due to photodissociation, especially on planets around G/K stars.
\rev{Thus, it is essential to use a detailed photochemical kinetics model to establish a comprehensive understanding on the relation between observable NH$_3$ and bulk nitrogen abundance.}

\rev{In this paper, we continue exploring the relation between observable nitrogen species and bulk nitrogen abundances\rerev{---the total nitrogen abundance in the atmosphere---}using a photochemical kinetics model.}
The organization of this paper is as follows.
\rev{In Section \ref{sec:N_map}, we briefly review the semi-analytical relationship of disequilibrium NH$_3$ and bulk nitrogen abundances predicted by \citetalias{Ohno&Fortney22a}.} \rev{In Section \ref{sec:photochemistry}, we apply a photochemical kinetics model to a wide range of planetary parameters to comprehensively understand the relation between observable NH$_3$ and bulk nitrogen abundance.} In Section \ref{sec:observation}, we investigate the observational feasibility of detecting nitrogen species in transmission and emission spectra.
In Section \ref{sec:discussion}, we discuss the effects of stellar spectral type, the presence of photochemical hazes, \rev{and day-night temperature contrast}.
In Section \ref{sec:summary}, we summarize our findings.

\section{\rev{Semi-analytical Relation Between NH$_3$ and Bulk Nitrogen Abundances}}\label{sec:N_map}
\begin{figure*}[t]
\centering
\includegraphics[clip, width=0.47\hsize]{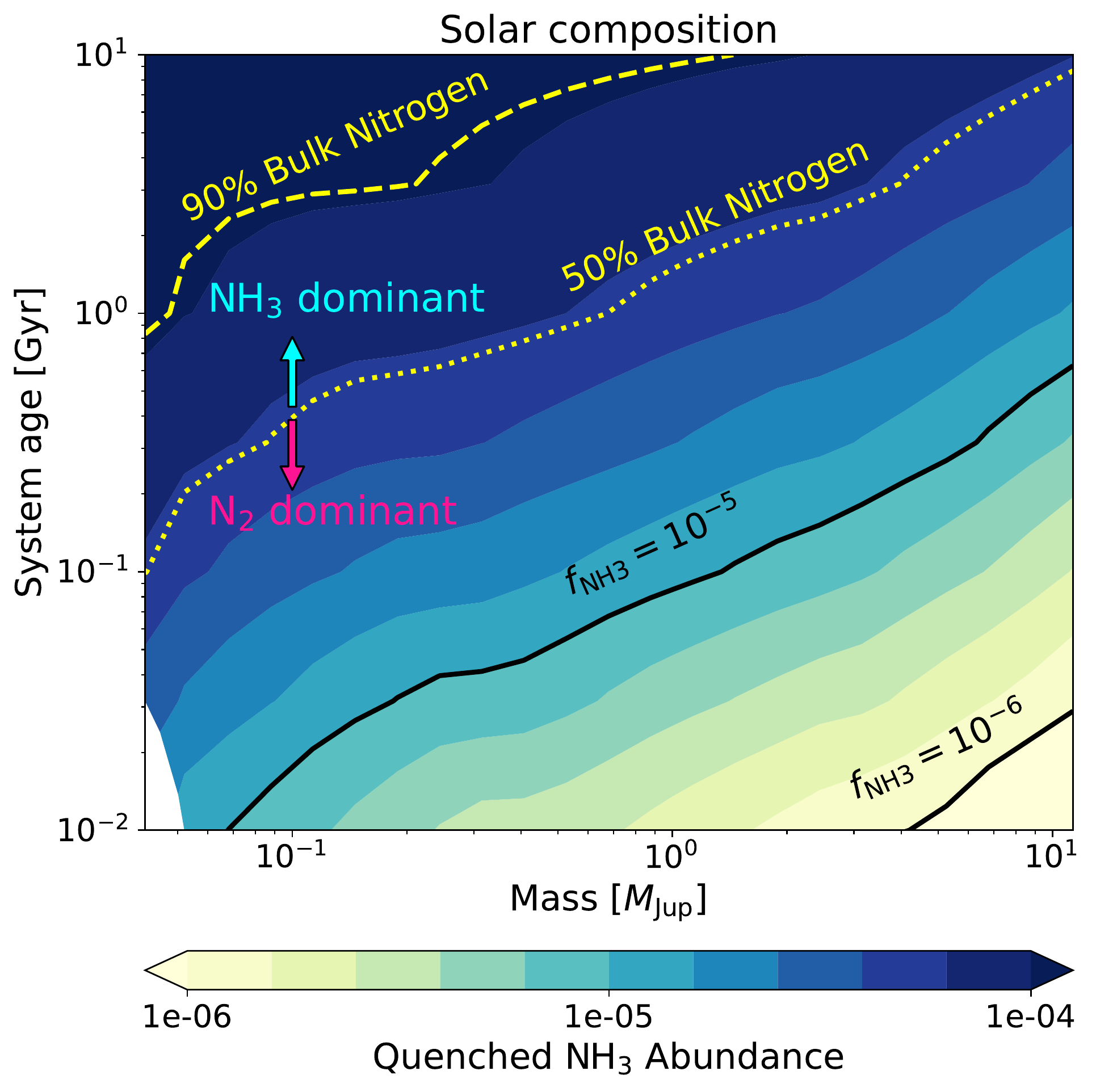}
\includegraphics[clip, width=0.47\hsize]{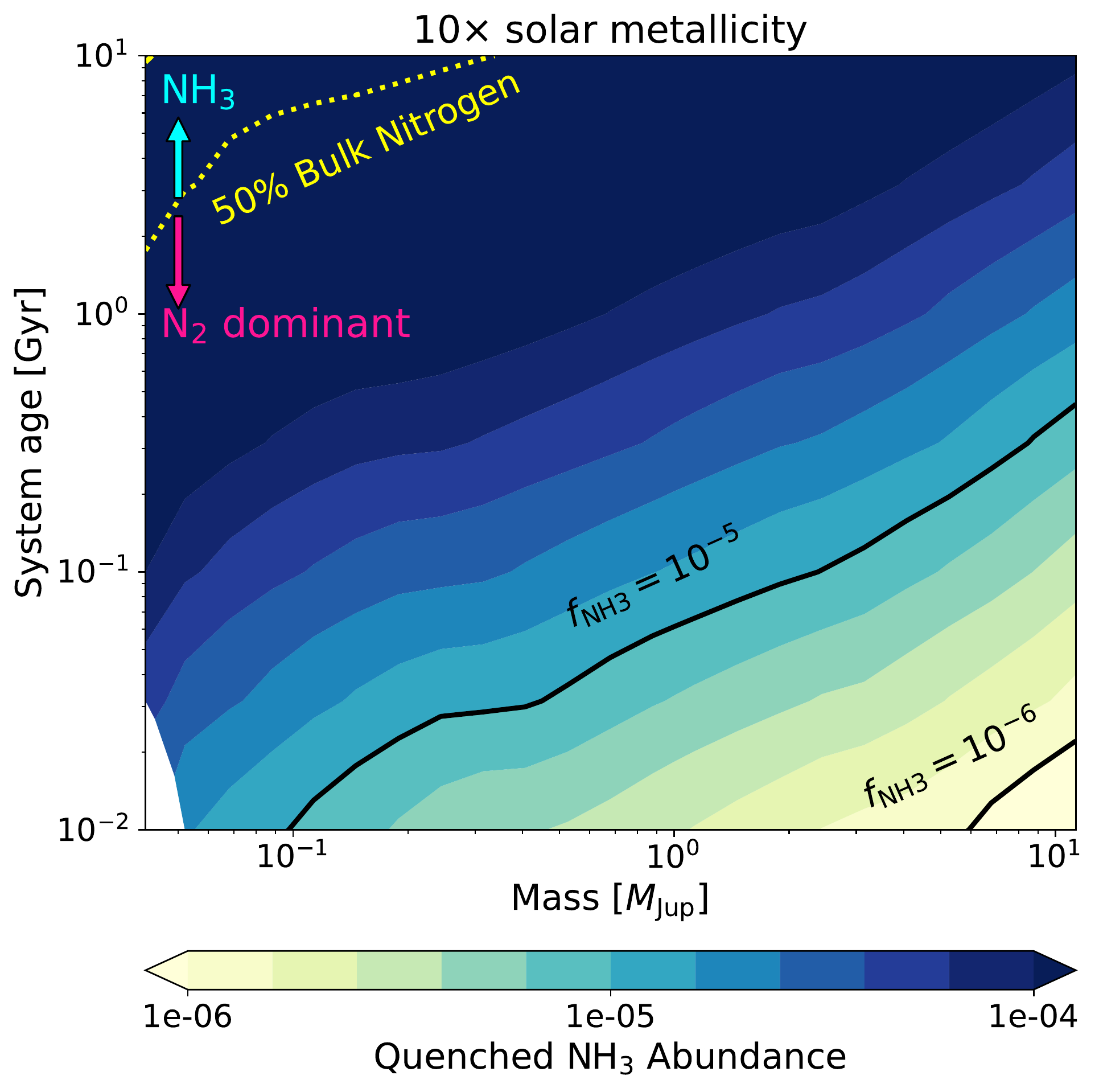}
\caption{\rev{The quenched NH$_3$ abundance predicted by the semi-analytic model of \citetalias{Ohno&Fortney22a} (Equation \ref{eq:NH3_analytic}) as a function of planetary mass and age, where $g$ and $T_{\rm int}$ are extracted from the thermal evolution tracks of \citet{Fortney+07}. 
The black line denotes the abundance contours of $f_{\rm NH_3}={10}^{-5}$ and ${10}^{-6}$, and yellow dashed and dotted lines show the contours corresponding to 90\% and 50\% of the bulk nitrogen abundance. The left and right columns show the results for solar metallicity and $10\times$ solar metallicity atmosphere, respectively, where we have assumed that the N/H ratio is scaled by the metallicity.}
}
\label{fig:NH3_map}
\end{figure*}
\rev{
In \citetalias{Ohno&Fortney22a}, based on the the law of mass action and a semi-analytic pressure-temperature (P-T) profile for deep adiabatic atmosphere, we established a semi-analytical model that predicts the vertically quenched NH$_3$ abundance of warm giant planets with $T_{\rm eq}\sim250$--$1200~{\rm K}$ as a function of the bulk nitrogen abundance.  This is given by
\begin{equation}\label{eq:NH3_analytic}
    \frac{f_{\rm NH3}}{f_{\rm N}}=\frac{\sqrt{1+8\mathcal{K}^{-1}}-1}{4}\mathcal{K},
\end{equation}
where $f_{\rm NH_{\rm 3}}$ is the volume mixing ratio of the quenched NH$_3$, $f_{\rm N}$ is the bulk nitrogen abundance, and $\mathcal{K}$ is the dimensionless parameter given by
\begin{eqnarray}\label{eq:K}
    \mathcal{K}&\approx& 3.46\times{10}^{-0.8{\rm [Fe/H]}}\left( \frac{f_{\rm N}}{10^{-4}}\right)^{-1}\left( \frac{g}{10~{\rm m~s^{-2}}}\right)^{4/3}\left( \frac{T_{\rm int}}{100~{\rm K}}\right)^{-16/3},
\end{eqnarray}
where [Fe/H] is the atmospheric metallicity, $g$ is the surface gravity, and $T_{\rm int}$ is the planetary intrinsic temperature. 
The bulk nitrogen abundance $f_{\rm N}$ is associated to nitrogen-to-hydrogen ratio N/H as
\begin{equation}\label{eq:N/H}
    f_{\rm N}={\rm \frac{N}{H_2+He } }=2f_{\rm H_2}{\rm N/H},
\end{equation}
where $f_{\rm H2}={\rm H_2/(H_2+He)}=0.859$ and $f_{\rm N}=1.16\times{10}^{-4}$ in the solar elemental abundances of \citet{Asplund+21}.
\citetalias{Ohno&Fortney22a} also provides an alternative form of Equation \eqref{eq:NH3_analytic}, which can be used to infer the bulk nitrogen abundance from a retrieved NH$_3$ abundance, as
\begin{eqnarray}\label{eq:N_diagnostic}
    \frac{f_{\rm N}}{f_{\rm NH_3}}&\approx&1 + 0.58\times{10}^{0.8{\rm [Fe/H]}}\left( \frac{f_{\rm NH_3}}{10^{-4}}\right)\left( \frac{g}{10~{\rm m~s^{-2}}}\right)^{-4/3}\left( \frac{T_{\rm int}}{100~{\rm K}}\right)^{16/3}.
\end{eqnarray}
This theory assumes that the quenched NH$_3$ abundance is insensitive to planetary equilibrium temperature and eddy diffusion coefficient.
This seemingly extreme assumption \rerev{is} valid because warm giant exoplanets have nearly the same deep atmosphere adiabatic profile at a given intrinsic flux and gravity (\citealt{Fortney+07,Fortney+20}; \citetalias{Ohno&Fortney22a}) and the thermochemical equilibrium abundance of NH$_3$ is nearly constant along the deep adiabat \rerev{for a given bulk nitrogen abundance} \citep{Saumon+06,Zahnle+14,Fortney+20}.
}

\rev{
Equation \eqref{eq:NH3_analytic} tells us how the quenched NH$_3$ abundance relates with the bulk nitrogen abundance.
For example, the NH$_3$ abundance is nearly the same as bulk nitrogen abundance, i.e., $f_{\rm NH_{\rm 3}}\approx f_{\rm N}$, at $\mathcal{K}\gg8$. 
By contrast, in the limit of low $\mathcal{K}$, the NH$_3$ abundance obeys $f_{\rm NH_3}\approx f_{\rm N}\sqrt{\mathcal{K}/2}$ that is lower than the bulk nitrogen abundance because of the conversion from NH$_3$ to N$_2$.
It is worth noting that the semi-analytic theory predicts that the quenched NH$_3$ abundance is insensitive to the metallicity in the N$_2$ rich regime (low $\mathcal{K}$ limit). 
When the bulk nitrogen abundance is scaled by the atmospheric metallicity, i.e., $f_{\rm N}\propto 10^{\rm[Fe/H]}$, Equation \eqref{eq:NH3_analytic} yields the nearly metallicity-independent NH$_3$ abundance of $f_{\rm NH_3}\propto 10^{0.1{\rm[Fe/H]}}$.
This insensitivity is owing to combined effects of hot deep interiors in the high metallicity atmospheres, and the preference for N$_2$ over NH$_3$ at higher metallicity \citep{Lodders&Fegley02,Moses+13b}. 
}

\rev{
Since the quenched NH$_3$ abundance depends only on $g$ and $T_{\rm int}$ for given metallicity, we can predict the quenched NH$_3$ abundance as a function of planetary mass and age, which control $g$ and $T_{\rm int}$.
Figure \ref{fig:NH3_map} shows the predicted quenched NH$_3$ abundance at various planetary masses and ages, taken from \citetalias{Ohno&Fortney22a}.
The quenched NH$_3$ abundance is higher at lower planetary mass and older age because of the cool deep atmosphere for these planets.
We predicted that, for solar composition atmospheres, the quenched NH$_3$ diagnoses more than 50\% of bulk nitrogen only when the planet has a sub-Jupiter mass ($\la1M_{\rm J}$) and old age ($\ga1~{\rm Gyr}$).
The discrepancy becomes even worse at high metallicity (for instance $\ga10$ solar) atmospheres, for which the quenched NH$_3$ abundance is significantly lower than the bulk nitrogen abundance at almost all values of planetary mass and age.
This analysis suggests the necessity of including a chemical model to properly infer the bulk nitrogen abundance from a retrieved NH$_3$ abundance, and the observable NH$_3$ abundance should be regarded as a lower limit of bulk nitrogen abundance in most cases. 
}

\rev{
The analysis of \citetalias{Ohno&Fortney22a} is based on a simple analytical argument.
In reality, thermochemistry and photochemistry can further alter the vertical distribution of NH$_3$ abundance.
In the next section, we use a photochemical kinetics model to better understand the relationship between observable NH$_3$ and the bulk nitrogen abundances.
}

\section{Comprehensive Investigations of Nitrogen Photochemistry}\label{sec:photochemistry}

\subsection{Numerical method}
\begin{table}[t]
  \caption{A grid of a part of thermal evolution tracks of giant planets with a core mass of $10 M_{\rm \oplus}$ in \citet{Fortney+07}, for which we apply VULCAN.}\label{table:1}
  \centering
  \begin{tabular}{r r r r r} \hline
     Mass $\rm[M_{\rm J}]$ & Radius [$R_{\rm J}$] & Age [Gyr]& Gravity $\rm[m~s^{-2}]$ & $T_{\rm int}$ $\rm[K]$\\ \hline \hline
    0.11 &1.14& 0.1 & 2.15 & 120.4 \\
    0.11 &0.89& 1 & 3.47 & 73.9\\
   0.11 &0.78& 10 & 4.60 & 39.0\\ 
    1.00 &1.21& 0.1 & 16.90 & 287.4 \\
    1.00 &1.11& 1 & 19.96 & 157.3\\
   1.00 &1.04& 10 & 22.71 & 83.7\\ \hline
  \end{tabular}
        \raggedright
\end{table}
\begin{figure*}[t]
\centering
\includegraphics[clip, width=0.49\hsize]{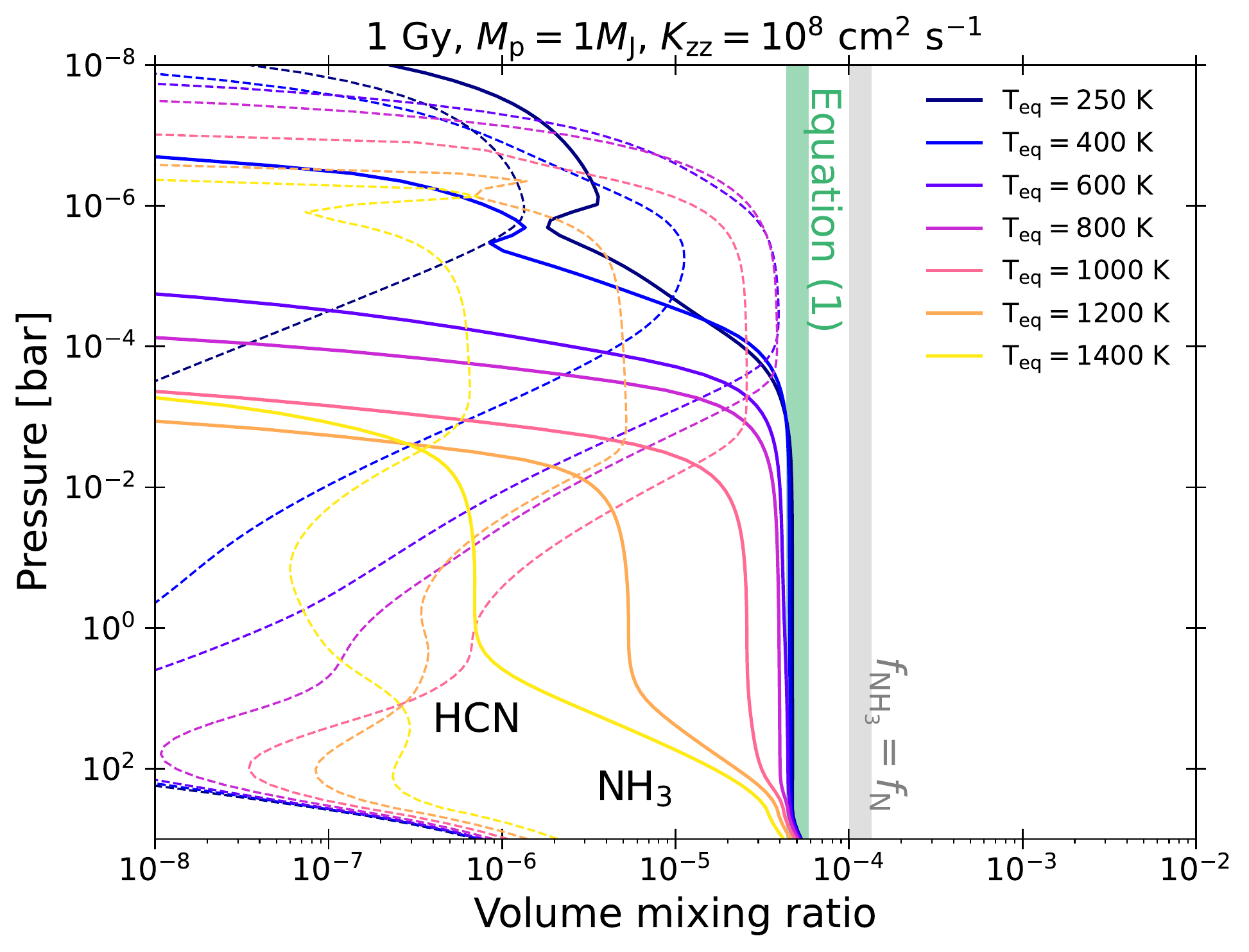}
\includegraphics[clip, width=0.49\hsize]{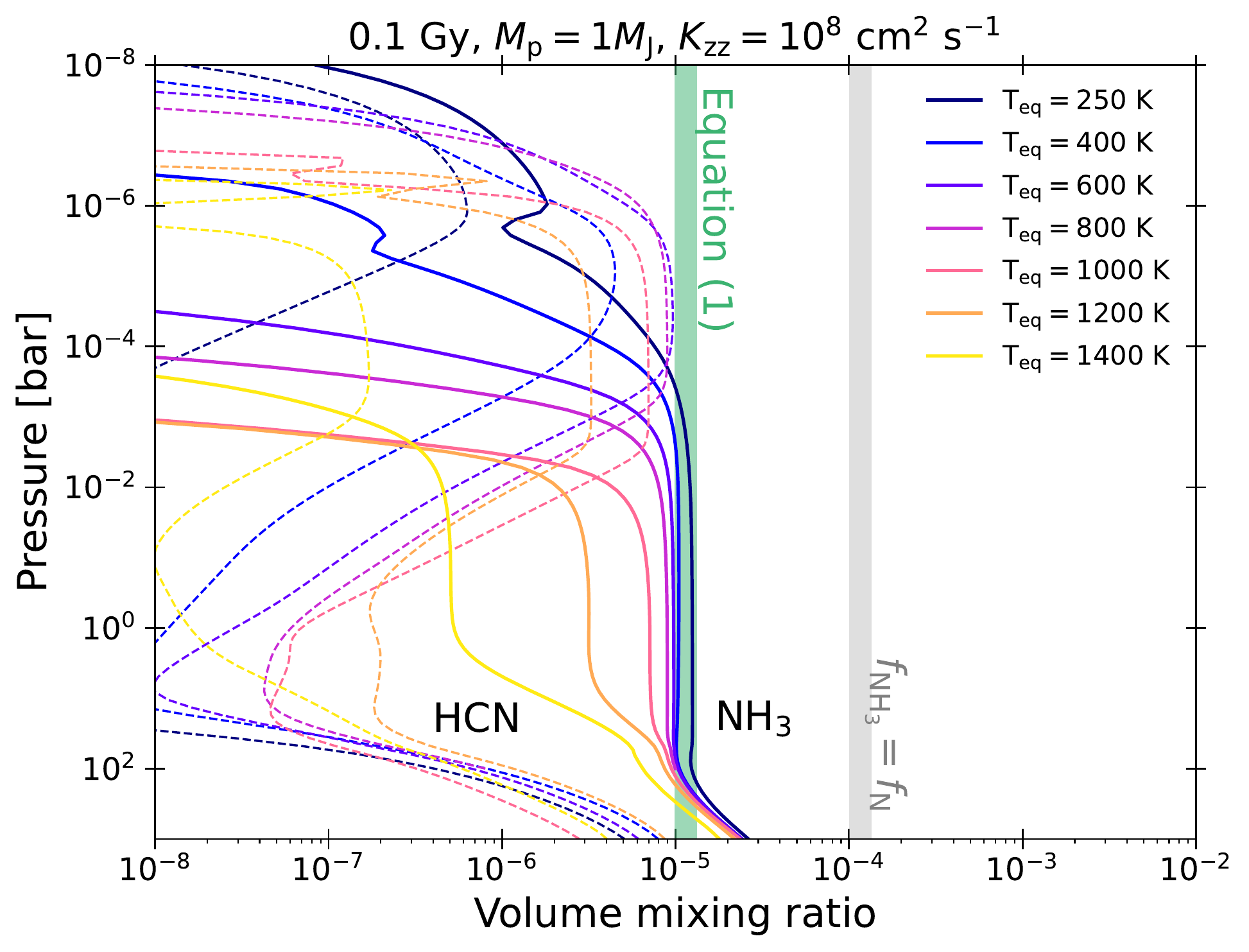}
\includegraphics[clip, width=0.49\hsize]{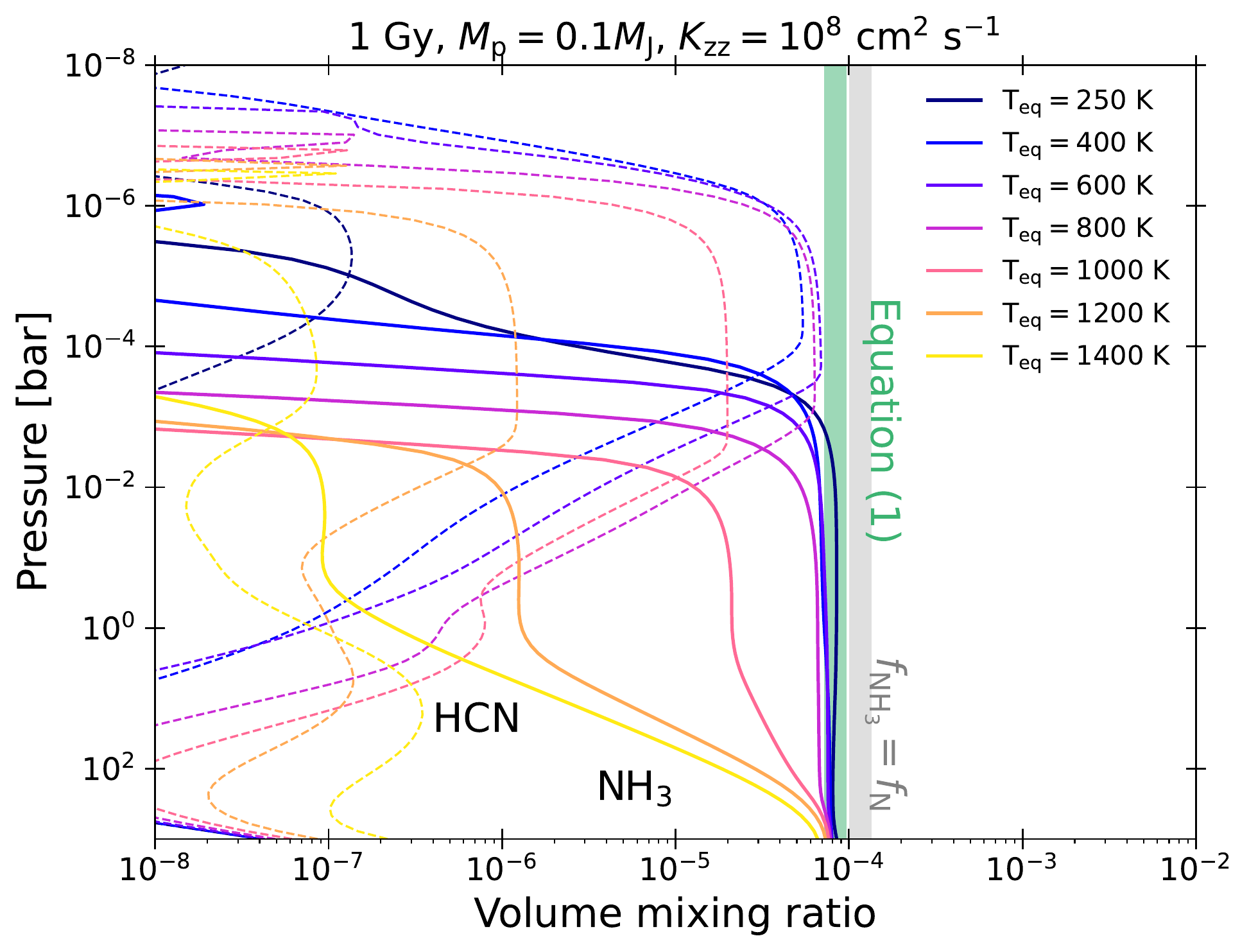}
\includegraphics[clip, width=0.49\hsize]{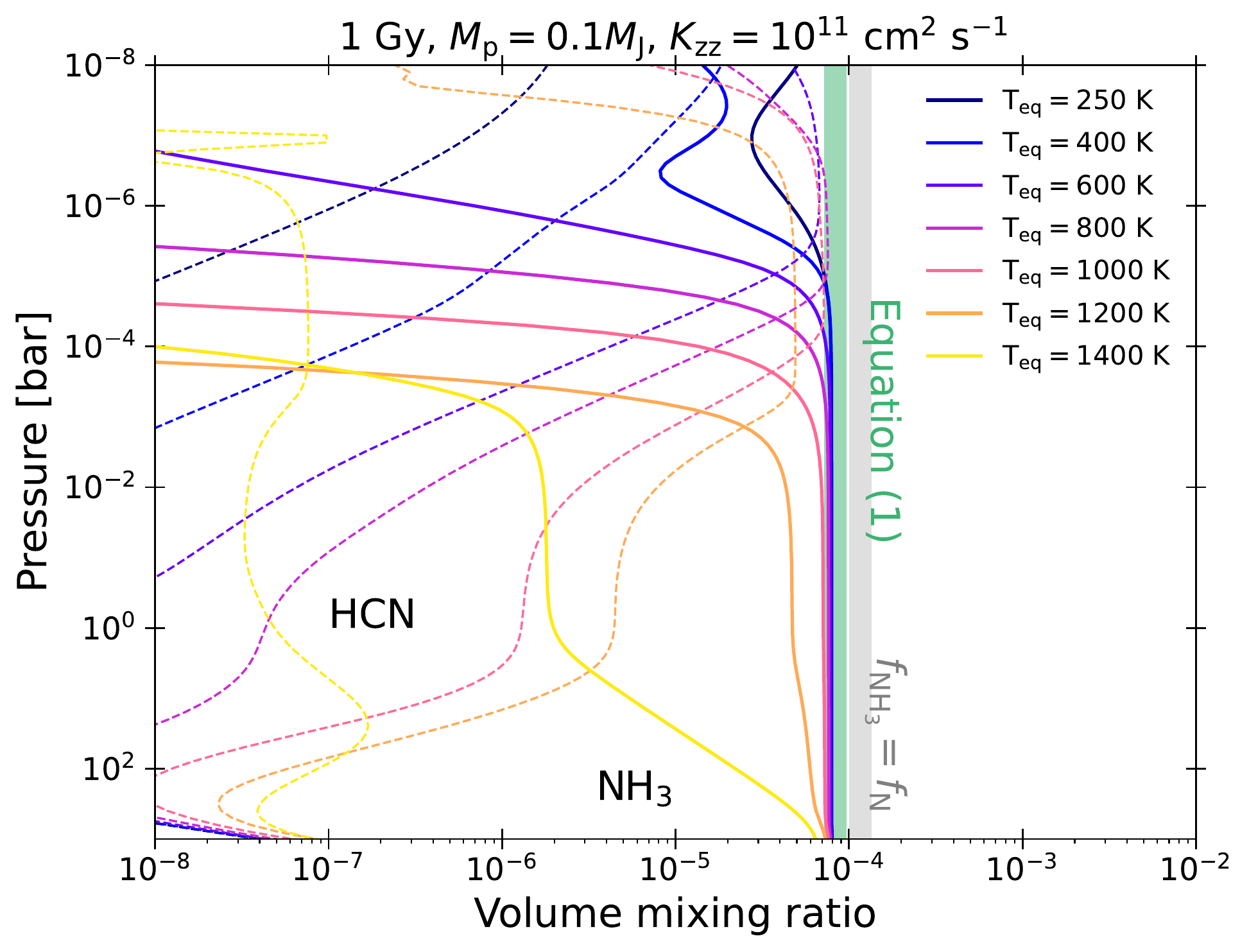}
\caption{Vertical distributions of NH$_3$ (solid lines) and HCN (dashed lines) volume mixing ratios. Different colored lines show the distributions for different planetary equilibrium temperature. The gray bold lines denote the abundance equal to the bulk nitrogen abundances, while the green bold lines denote the NH$_3$ abundance predicted by our semi-analytical model (Equation \ref{eq:NH3_analytic}). We have assumed a solar composition atmosphere in this figure.
\rerev{The left top panel shows the results for a $1~{\rm Gy}$ old Jupiter mass planet with $K_{\rm zz}={10}^{8}~{\rm {cm}^2~s^{-1}}$, the right top panel instead sets the age of $0.1~{\rm Gy}$ to demonstrate the effect of planetary age, the left bottom panel sets $M_{\rm p}=0.1M_{\rm J}$ to demonstrate the planetary mass dependence, and right bottom panel sets $K_{\rm zz}={10}^{11}~{\rm cm^2~s^{-1}}$ to be compared with left bottom panel for showing the $K_{\rm zz}$ dependence.}
}
\label{fig:VULCAN_result}
\end{figure*}
\begin{figure}[t]
\centering
\includegraphics[clip, width=\hsize]{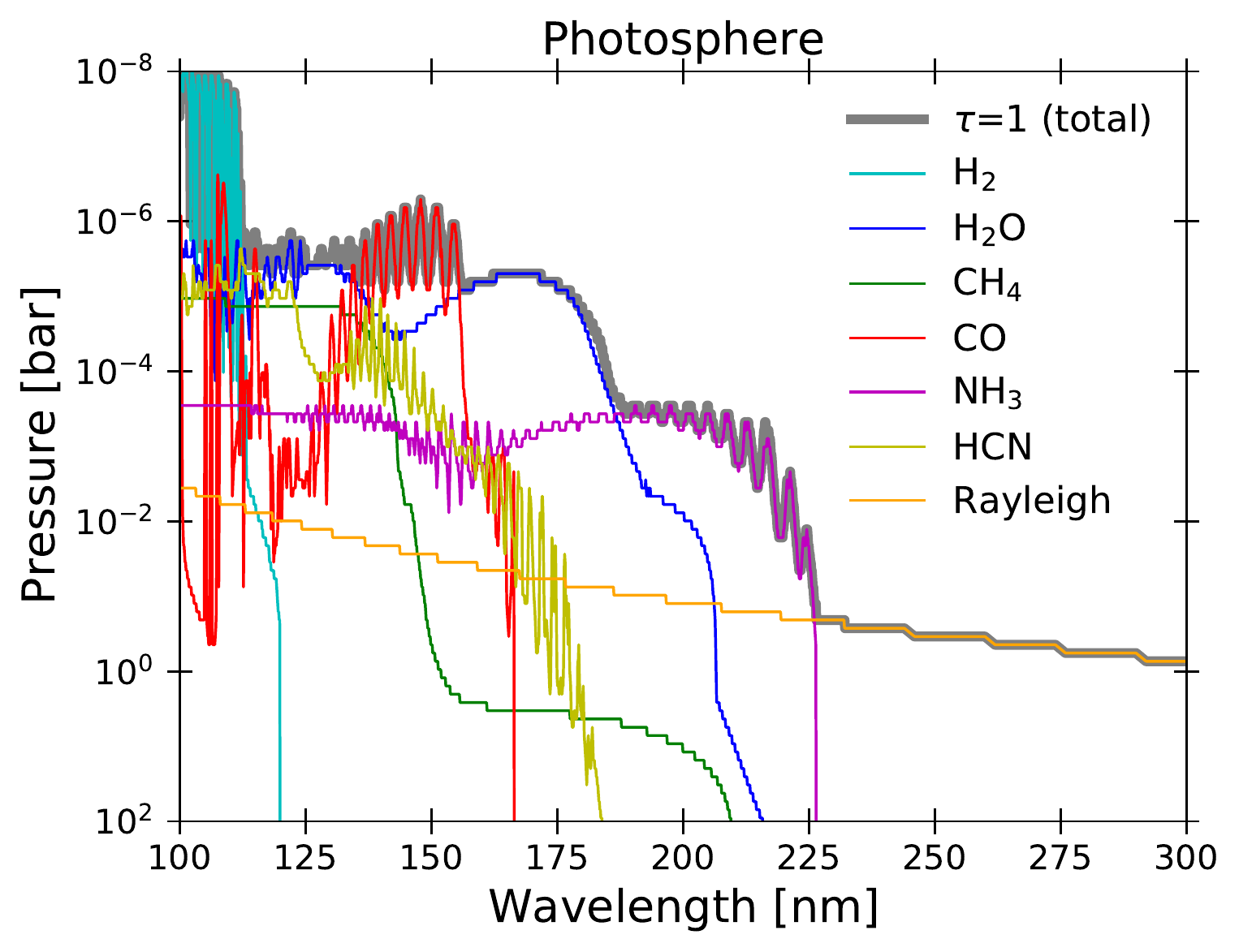}
\caption{Photospheric pressure level of at each wavelength across the UV. The thick gray line shows the pressure level of vertical optical depth of unity, and other colored lines show the contribution of each molecule.
We assume a solar composition atmosphere, $M_{\rm p}=1M_{\rm J}$, age of $1~{\rm Gyr}$, $T_{\rm eq}=800~{\rm K}$, and $K_{\rm zz}={10}^{8}~{\rm {cm}^2s^{-1}}$.
}
\label{fig:photosphere}
\end{figure}
\begin{figure*}[t]
\centering
\includegraphics[clip, width=\hsize]{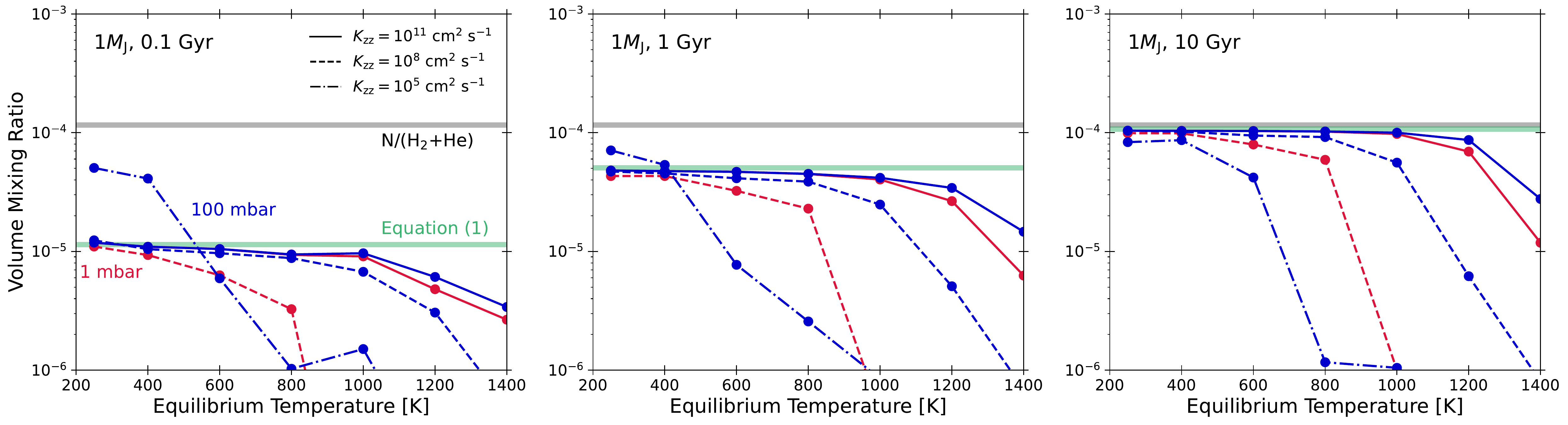}
\includegraphics[clip, width=\hsize]{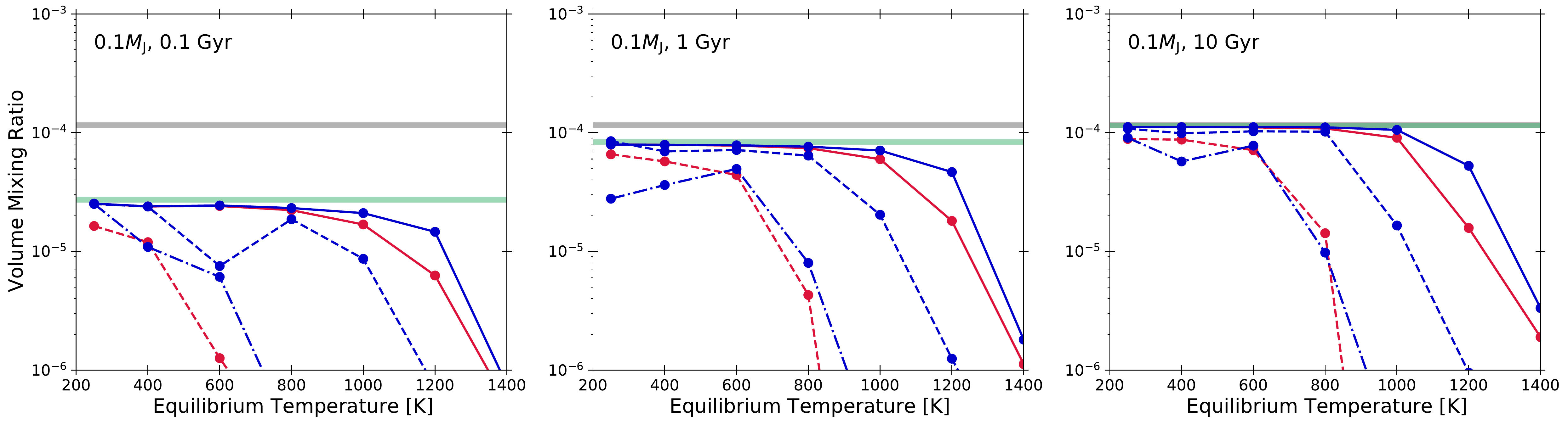}
\caption{NH$_3$ abundance at pressure levels of $1$ mbar (red lines) and $100~{\rm mbar}$ (blue lines). The solid, dashed, and dash-dot lines show the abundances for $K_{\rm zz}={10}^{11}$, $10^{8}$, and ${10}^{5}~{\rm {cm}^2~s^{-1}}$, respectively. From left to right, each panel show the abundances for planets at the age of $0.1$, $1$, and $10~{\rm Gyr}$. The top and bottom panels show the results for Jupiter-mass ($1M_{\rm J}$) and Neptune-mass ($0.11M_{\rm J}$) planets, respectively.
The gray lines denote the bulk nitrogen abundance, and the green lines denote the quenched nitrogen abundance predicted by Equation \ref{eq:NH3_analytic}.
We have assumed solar composition atmospheres.
}
\label{fig:NH3_mbar}
\end{figure*}

\begin{figure*}[t]
\centering
\includegraphics[clip, width=0.49\hsize]{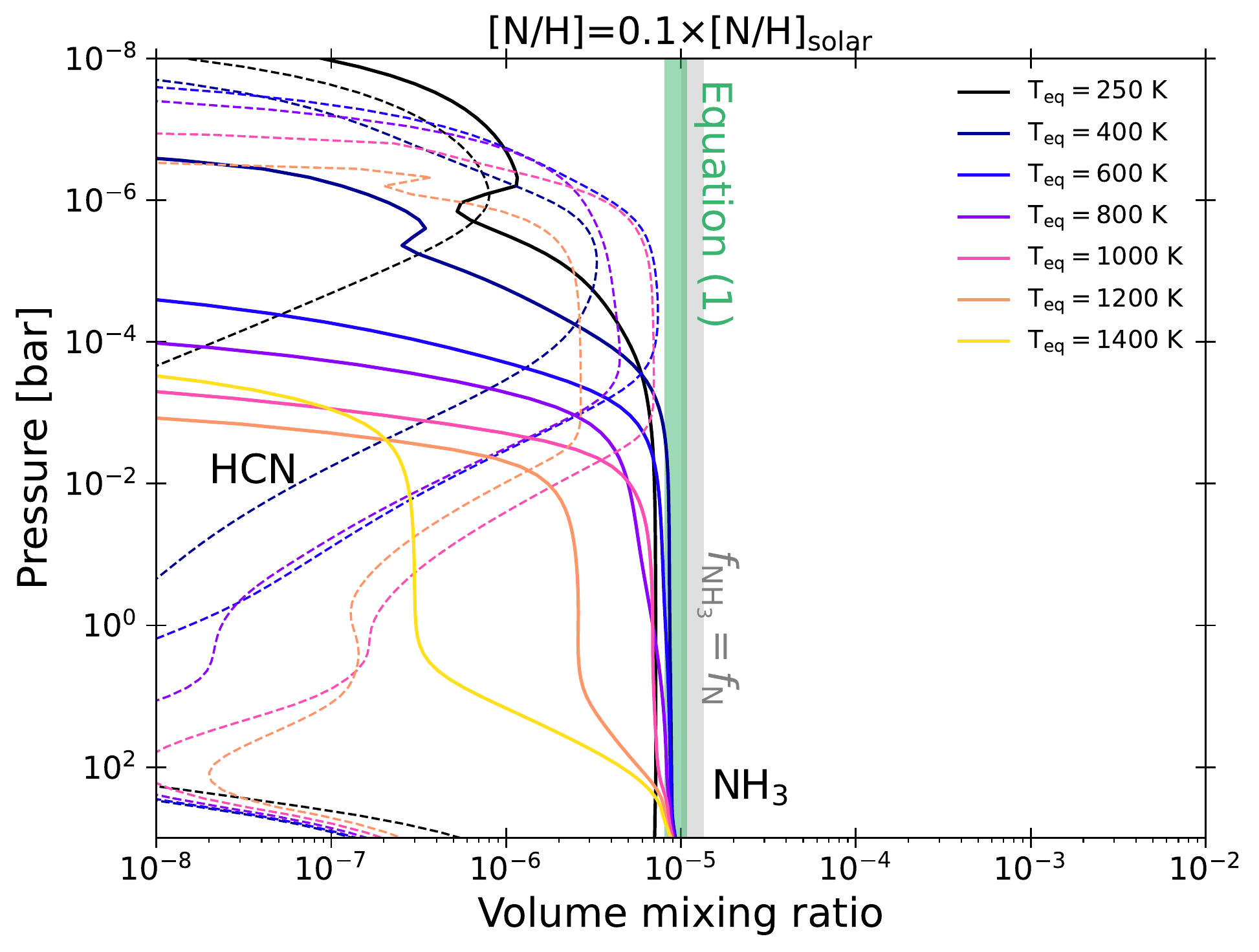}
\includegraphics[clip, width=0.49\hsize]{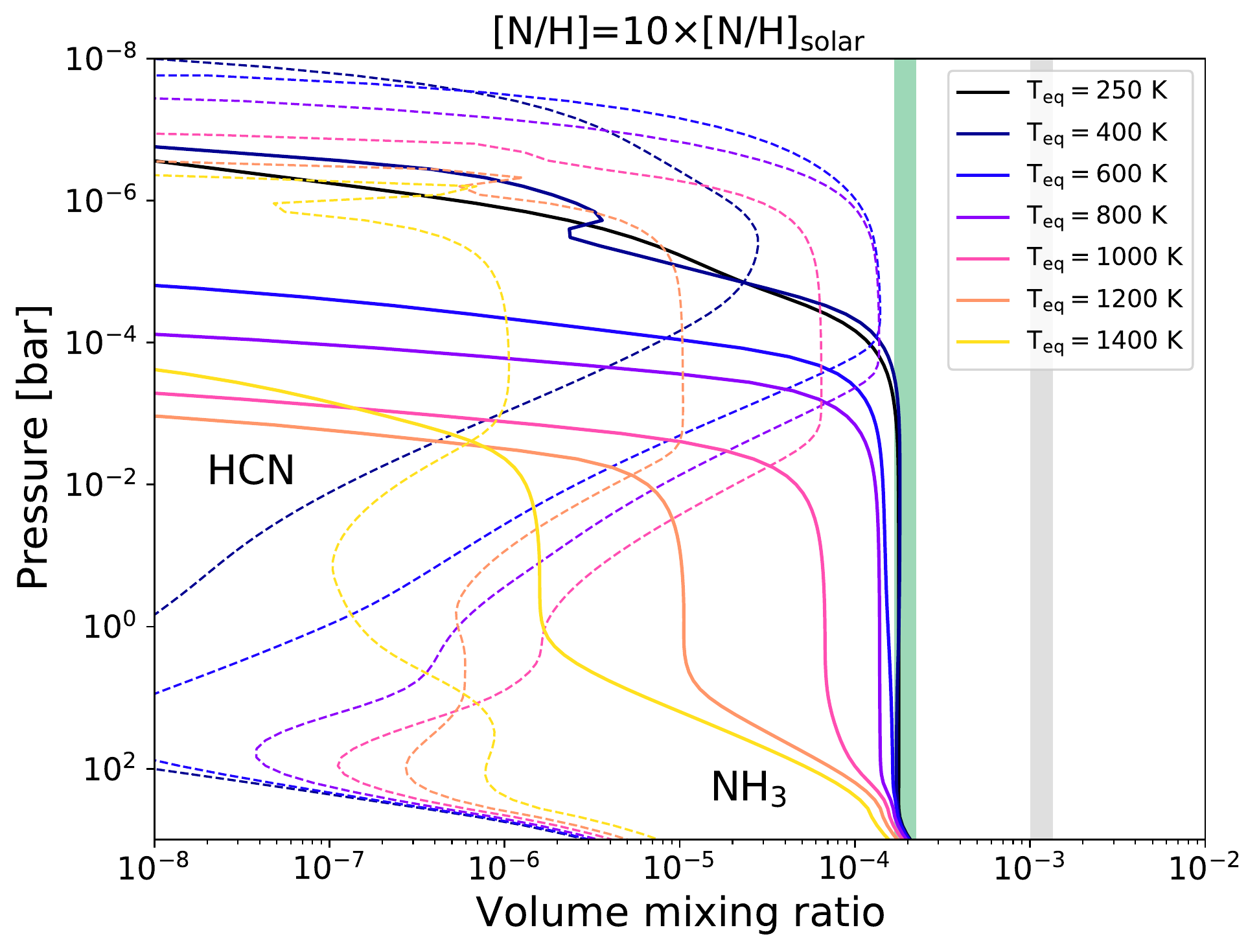}
\includegraphics[clip, width=0.49\hsize]{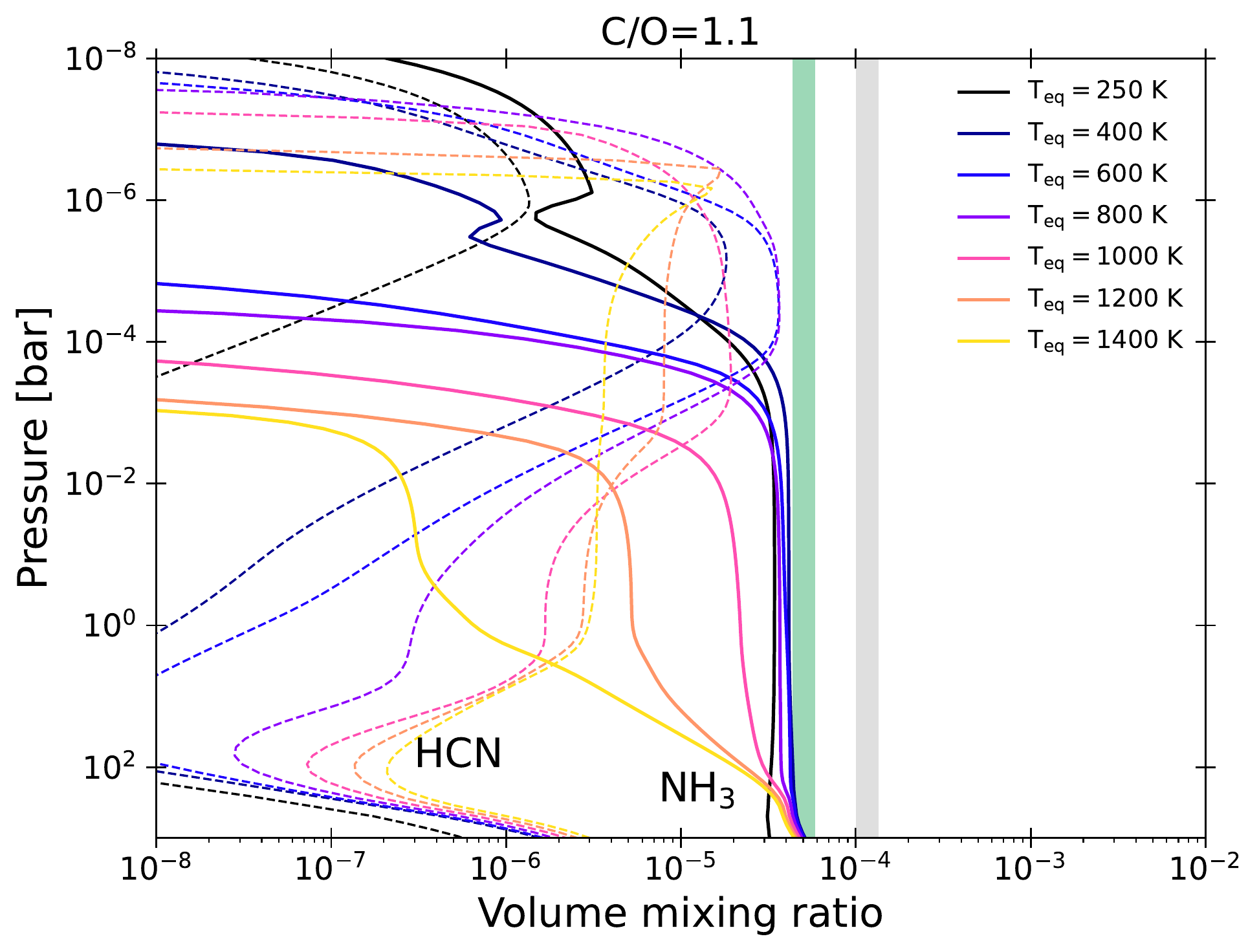}
\includegraphics[clip, width=0.49\hsize]{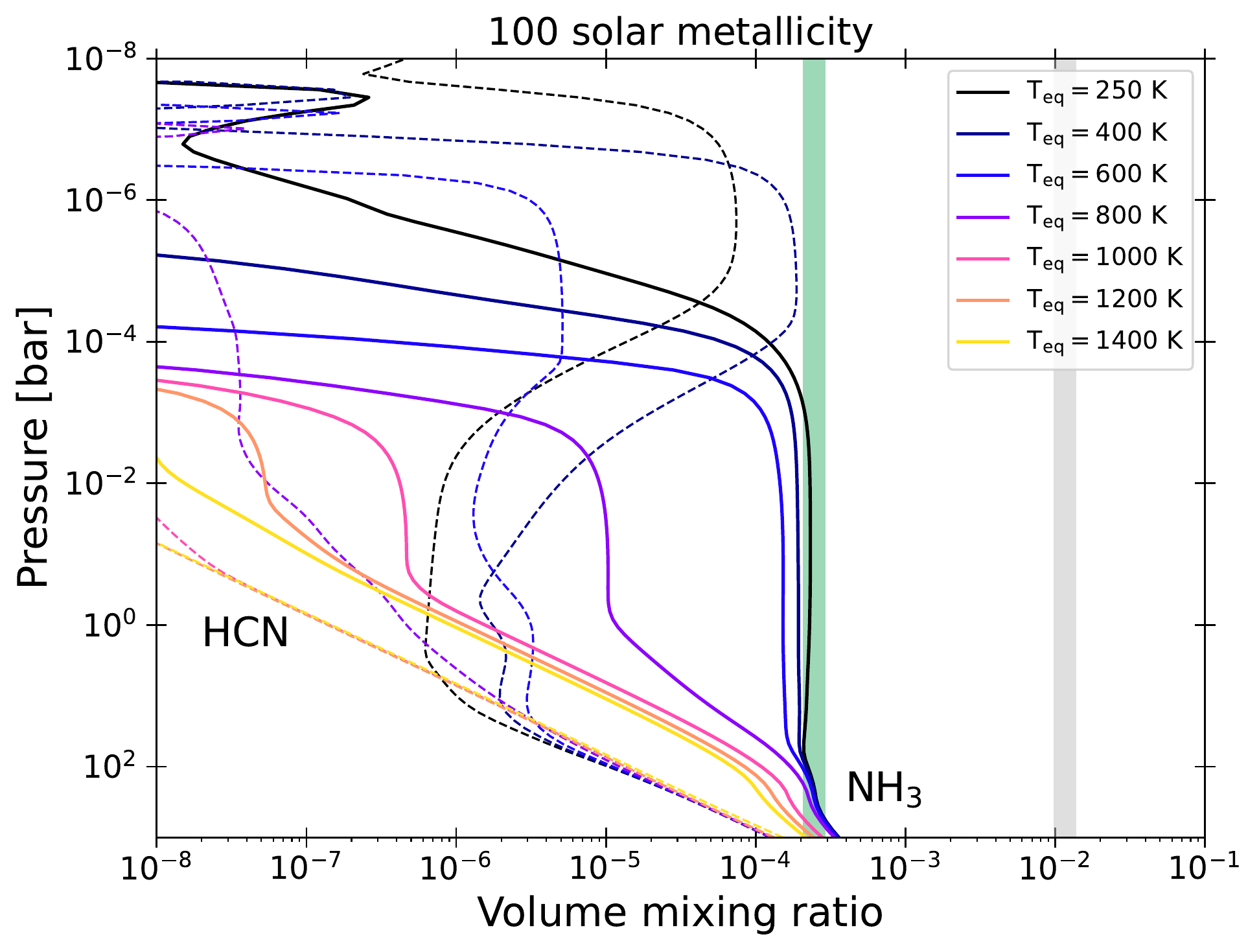}
\caption{Same as Figure \ref{fig:VULCAN_result}, but for different atmospheric compositions.  The upper left and right panels show the NH$_3$ (solid curves) and HCN (dashed curves) distributions for the N/H ratio of $0.1 \times$ and $10 \times$ solar values, the lower left panel show the results for C/O=1.1, and the lower right panel shows the results for the atmospheric metallicity of 100$\times$ the solar value. We assume a Jupiter-mass planet at the age of $1~{\rm Gyr}$ for the variation of N/H and C/O, while we assume a Neptune-mass planet for the 100$\times$ solar atmospheres. We set $K_{\rm zz}={10}^{8}~{\rm {cm}^2~s^{-1}}$ for all calculations.
}
\label{fig:VULCAN_result2}
\end{figure*}

\begin{figure}[t]
\centering
\includegraphics[clip, width=\hsize]{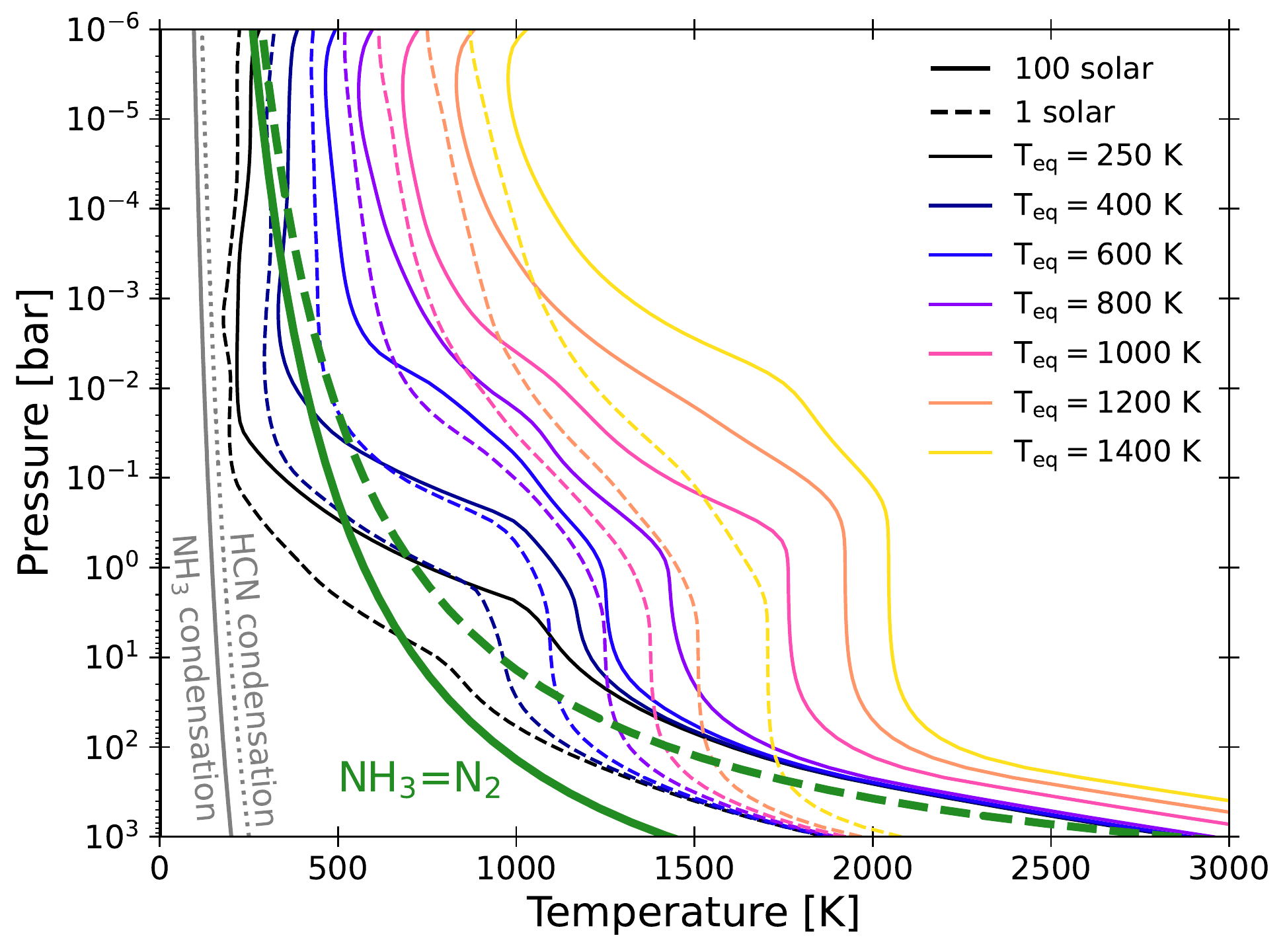}
\includegraphics[clip, width=\hsize]{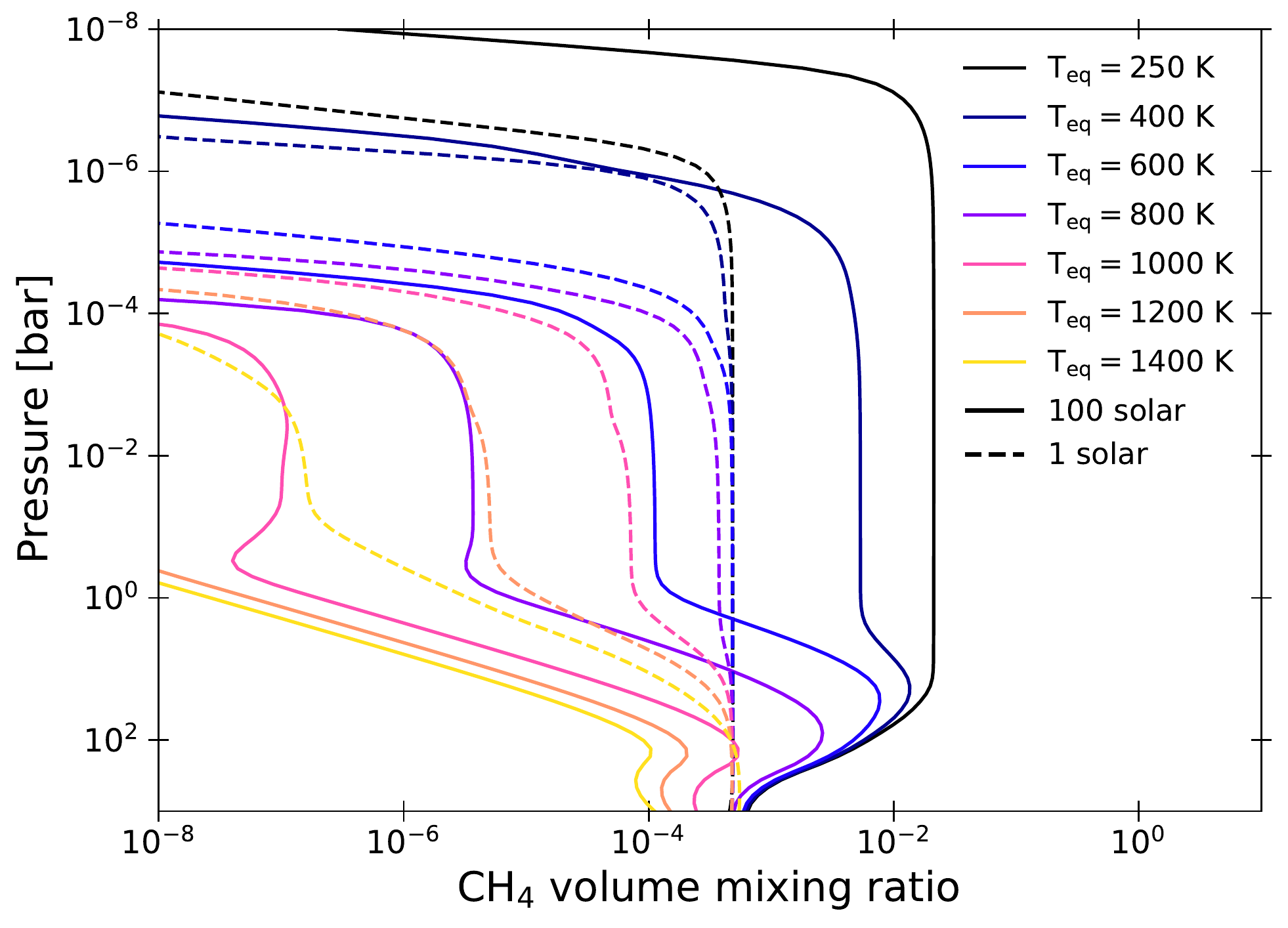}
\caption{Atmospheric \emph{P--T} profiles (top panel) and vertical distributions of CH$_4$ abundances (bottom panel) in a Neptune-mass planet ($0.11M_{\rm J}$) with the age of $1~{\rm Gyr}$. Different colored lines show the profiles for different equilibrium temperature. The solid and dashed lines show the profiles for 100$\times$ solar metallicity and solar composition atmospheres, respectively. 
The green dashed and solid lines in the top panel denote the equal equilibrium abundance \emph{P--T} relation of N$_2$ and NH$_3$, computed by the law of mass action as described in \citet{Zahnle+14} for $\rm NH_3=N_2$, for solar and 100$\times$ solar metallicity atmospheres, respectively.
\rerev{The gray solid and dotted lines in the top panel also denote the condensation temperature of NH$_3$ and HCN when assuming the volume mixing ratio of ${10}^{-4}$, typical peak abundances seen in our photochemical calculations (see Figures \ref{fig:VULCAN_result} and  \ref{fig:VULCAN_result2}), where we have used the vapor pressure of \citet{Fray&Schmitt09}.}
}
\label{fig:100solar}
\end{figure}

Here, we numerically investigate atmospheric chemical compositions using a publicly available photochemical kinetics code, VULCAN \citep{Tsai+17,Tsai+21}.
The code computes the steady state vertical distributions of chemical compositions by solving the transport equations given by
\begin{equation}
    \frac{\partial n_{\rm i}}{\partial t}=\mathcal{P}_{\rm i}-\mathcal{L}_{\rm i}-\frac{\partial \Phi_{\rm i}}{\partial z},
\end{equation}
where $z$ is the altitude, $n_{\rm i}$ is the number density of species i, $\mathcal{P}_{\rm i}$ and $\mathcal{L}_{\rm i}$ are their production and loss rates, and $\phi_{\rm i}$ is the vertical flux due to eddy and molecular diffusion. 
While several theoretical studies investigated the eddy diffusion coefficient $K_{\rm zz}$ on exoplanets using global circulation models \citep{Parmentier+13,Charnay+15,Zhang&Showman18a,Zhang&Showman18b,Menou19,Komacek+19,Menou21,Tan22}, it remains poorly constrained from observations \citep[see][]{Kawashima&Min21}.
Thus, we test a broad range of eddy diffusion coefficient, $K_{\rm zz}={10}^{5}$, ${10}^{8}$, and ${10}^{11}~{\rm {cm}^2~s^{-1}}$, for the sensitivity tests.
We assume a zero flux condition for both upper and lower boundaries. 
We set the depth of the lower boundary so that thermochemical equilibrium is maintained there (typically $1000~{\rm bar}$, depending on the thermal state of the deep atmosphere and $K_{\rm zz}$).
We adopt the C-H-N-O chemistry network implemented in VULCAN as a default.
We assume the solar photospheric elemental abundances of \citet{Asplund+21} for fiducial simulations.
To explore a range of parameter space for planetary properties shown in Figure \ref{fig:NH3_map}, we extract the surface gravity and intrinsic temperature from the thermal evolution tracks of \citet{Fortney+07}\footnote{Grids of the evolution tracks are available at \url{https://www.ucolick.org/~jfortney/models.htm}} for several representative planetary masses and ages, as summarized in Table \ref{table:1}. 
Then, we compute the atmospheric \emph{P--T} profiles using the ``EGP'' radiative-convective equilibrium model extensively used in previous studies \footnote{A Python version of the model has recently been made publicly available \citep{Mukherjee+22a}.} (e.g., \citealt[][]{McKay+89,Marley&McKay99,Fortney+05,Fortney+07,Fortney+08,Morley+12,Marley&Robinson15,Thorngren+19,Gao+20}; \citetalias{Ohno&Fortney22a}) for various planetary equilibrium temperatures.
Our fiducial simulations assume the solar spectrum \citep{Gueymard18}, \rerev{where we set the orbital distance so that the planetary equilibrium temperature (for zero Bond albedo with full heat redistribution) coincides with a specified value.}
We test the effects of different stellar spectra in Section \ref{sec:discussion_stellar}.
\rerev{We neglect the condensation of gas molecules since the planets studied this paper are too warm to cause condensation of NH$_3$ and HCN (see top panel of Figure \ref{fig:100solar}).}

\subsection{Effects of Planetary Mass, Age, and Eddy Diffusion}\label{sec:result_age}
We first explore how the NH$_3$ and HCN abundances vary with various planetary properties.
Figure \ref{fig:VULCAN_result} shows the vertical distributions of NH$_3$ and HCN across equilibrium temperature, planetary mass, age, and eddy diffusion coefficient.
In general, the NH$_3$ abundance is nearly constant in the middle atmosphere owing to vertical mixing, while it decreases with increasing altitudes in the upper atmosphere due to  photodissociation.
This general trend is in agreement with previous studies \citep[e.g.,][]{Moses+11,Line+11,Miller-Ricci+12,Kawashima&Ikoma18}.
As predicted in Section \ref{sec:N_map}, the quenched NH$_3$ abundances in middle atmospheres are nearly independent of the equilibrium temperature at $T_{\rm eq}\la 800~{\rm K}$.
For $T_{\rm eq}\ga 1000~{\rm K}$, the NH$_3$ abundance gradually decreases with decreasing pressure at $P\sim 1$--$100~{\rm bar}$ because of the efficient thermochemical conversion to N$_2$ due to high temperatures. 
This trend is consistent with previous studies of hot Jupiters \citep{Moses+11}.

The ratio of the quenched NH$_3$ to bulk nitrogen abundance depends on planetary mass and age.
The gray bold lines in Figure \ref{fig:VULCAN_result} denote the bulk nitrogen abundance $f_{\rm N}$.
For Jupiter-mass planets ($M_{\rm p}=1M_{\rm J}$) at $1~{\rm Gyr}$ age  (left top panel of Figure \ref{fig:VULCAN_result}), the quenched NH$_3$ abundance is lower than the bulk nitrogen abundance by a factor of $\sim2$. 
The depletion factor even reaches a factor of $\sim10$ for an age of $\sim0.1{\rm Gyr}$  (right top panel of Figure \ref{fig:VULCAN_result}).
On the other \rerev{hand}, the NH$_3$ abundance is almost equal to the bulk nitrogen abundances for Neptune mass planets ($M_{\rm p}=0.11M_{\rm J}$) at $1~{\rm Gyr}$ (left bottom panel of Figure \ref{fig:VULCAN_result}).
These results are consistent with our prediction made in the previous section (Figure \ref{fig:NH3_map}).
Thus, higher planetary mass and younger age do act to deplete the quenched NH$_3$ as compared to the bulk nitrogen abundance.

It is worth noting that our semi-analytical prediction of the quenched NH$_3$ abundance (Equation \ref{eq:NH3_analytic}) matches the numerical results quite well. 
The green bold lines in Figure \ref{fig:VULCAN_result} show the NH$_3$ abundance given by Equation \eqref{eq:NH3_analytic}, which coincides with the quenched NH$_3$ abundance below the NH$_3$ photodissociation base.
Thus, even if the planetary mass and age fall into the regime where NH$_3$ depletion is expected, one may still be able to use our semi-analytic diagnostic model (Equation \ref{eq:N_diagnostic}) to give a first-order estimate on the bulk nitrogen abundance from the NH$_3$ abundance. 

Of course photochemistry may substantially deplete the observable NH$_3$ abundances at low pressure.
As seen in the results for an eddy diffusion coefficient of $K_{\rm zz}={10}^8~{\rm {cm}^2~s^{-1}}$ (upper panels and lower left panel of Figure \ref{fig:VULCAN_result}), the NH$_3$ abundance drops off at $P\la{10}^{-3}$--${10}^{-4}~{\rm bar}$ through photodissociation. 
This pressure level is significantly deeper than the pressure level where the many other chemical species experience photodissociation, say $\sim {10}^{-5}$--${10}^{-6}~{\rm bar}$.
This difference is because NH$_3$ is fragile to UV photons \rerev{to} relatively longer wavelengths.
To demonstrate this, Figure \ref{fig:photosphere} shows the photospheric pressure level at UV wavelengths.
While UV photons at $\la 190~{\rm nm}$ are mostly absorbed by other chemical species, such as H$_2$O, at $\sim {10}^{-5}$--${10}^{-6}~{\rm bar}$, near-UV photons at $190$--$220~{\rm nm}$ penetrate to deeper atmospheres because of negligible photolysis cross sections of other molecules.
Those near-UV photons largely act to deplete NH$_3$ at $\sim10^{-3}~{\rm bar}$, which explains why NH$_3$ is vulnerable to photochemistry at relatively higher pressures. 
We refer readers to \citet{Hu21} for extended discussions on the depletion of NH$_3$ through photodissociation in middle atmosphere regions.
We note that our simulations do not account for the presence of aerosols, such as photochemical hazes, while it may act to prevent UV photons from penetrating to deep atmospheres, as discussed in Section \ref{sec:discussion_haze}.

The strength of eddy diffusion affects where the photochemical depletion of NH$_3$ takes place.
The left and right bottom panels of Figure \ref{fig:VULCAN_result} show the results for $M_{\rm p}=0.11M_{\rm J}$ and $1~{\rm Gyr}$ but for $K_{\rm zz}={10}^{8}$ and $K_{\rm zz}={10}^{11}~{\rm {cm}^2~s^{-1}}$.
Higher $K_{\rm zz}$ results in the quenched NH$_3$ abundance continuing to higher altitudes, as the fast eddy diffusion can more easily compensate against the NH$_3$ loss by photodissociation (see also \citealt{Hu21}).
On the other hand, different values of $K_{\rm zz}$ have negligible impacts on the quenched abundance of NH$_3$, in agreement with previous studies \citep{Zahnle+14,Fortney+20}.

Once the photolysis of NH$_3$ sets in, most of NH$_3$ is converted to HCN in the upper atmosphere for the equilibrium temperature of $T_{\rm eq}=400$--$1400~{\rm K}$.
In Figure \ref{fig:VULCAN_result}, one can notice that HCN abundances at upper atmospheres, say $P\la {10}^{-3}$--${10}^{-4}~{\rm bar}$, are approximately the same as the NH$_3$ abundances below the photodissociation base.
\rev{This result indicates that the photodissociated NH$_3$ is almost completely converted to HCN. Thus, one might be able to evaluate the bulk nitrogen abundance from the measurement of HCN abundance even if photodissociation significantly depletes NH$_3$ at pressure levels probed by observations. If the abundance profiles of NH$_3$ and HCN are constrained simultaneously, it would provide complementary information to infer the bulk nitrogen abundance.}

Although the detailed analysis of atmospheric chemistry is beyond the scope of this paper, we here briefly introduce HCN synthesis processes.
HCN can be produced through photodissociation of NH$_3$ under the presence of CH$_4$.
For example, \citet{Moses+11} and \citet{Line+11} identified the following channel of HCN production:
\begin{eqnarray}
\nonumber
\mathrm{ CH_4 + H} &\longrightarrow& \mathrm{CH_3 + H_2}\\
\nonumber
\mathrm{ NH_3} + h\nu &\longrightarrow& \mathrm{NH_2 + H}\\
\nonumber
\mathrm{ NH_2 + H} &\longrightarrow& \mathrm{ NH + H_2}\\
\nonumber
\mathrm{ NH + H }&\longrightarrow& \mathrm{N + H_2}\\
\nonumber
\mathrm{ N + CH_3} &\longrightarrow& \mathrm{H_2CN + H}\\
\nonumber
\mathrm{ H_2CN+H} &\longrightarrow& \mathrm{HCN + H_2}\\
\hline
\nonumber
\\
\nonumber
\text{Net:}~\mathrm{ NH_3 + CH_4} &\longrightarrow& \mathrm{ HCN + 3H_2}
\end{eqnarray}
HCN synthesis initiating from NH$_3$ photodissociation requires CH$_3$ produced through the reaction of CH$_4$ with H, where H can be provided by the photodissociation of NH$_3$.
In addition to the above channel through H$_2$CN, HCN can also directly form from N and CH$_3$ as \citep[][]{Kawashima&Ikoma18,Hobbs+19}
\begin{eqnarray}
\nonumber
\mathrm{ N + CH_3} &\longrightarrow& \mathrm{H_2 + HCN}
\end{eqnarray}
On the other hand, for temperate to cold planets, \citet{Hu21} identified the following channels without bypassing NH:
\begin{eqnarray}
\nonumber
\mathrm{ NH_2 + CH_3} &\longrightarrow& \mathrm{ CH_5N}\\
\nonumber
\mathrm{ CH_5N }+h\nu&\longrightarrow& \mathrm{HCN + 2H_2}
\end{eqnarray}
\rerev{though CH$_5$N photodissociation does not actually directly yield HCN but yield CH$_3$NH that subsequently yield CH$_2$NH and HCN via reaction with atomic H and photodissociation \citep[see Section 3.2 of][]{Moses+10}.
As seen those examples, the major reaction pathway of HCN production may depend on specific photochemical model used.
}
HCN is destroyed either by the reaction of HCN $+$ H $\rightarrow$ CN $+$ H$_2$ or photodissociation of HCN $+ h\nu\rightarrow$ CN $+$ H, but CN quickly reacts with H$_2$ to reproduce HCN again \citep{Moses+11}.
In our calculations, the H$_2$CN $+$ H channel dominates over the N + CH$_3$ channel. 
The relative importance of the CH$_5$N channel in warm exoplanets remains unclear since the default chemical network of the VULCAN does not involve reactions around CH$_5$N \citep[see][]{Tsai+21}.
Nevertheless, we predict that the addition of the CH$_5$N channel would barely affect our results in terms of HCN vertical distributions.
This is because the HCN production rate is eventually limited by the upward flux of NH$_3$ in the limit of fast HCN synthesis, and complete conversion of NH$_3$ to HCN in Figure \ref{fig:VULCAN_result} implies that the HCN synthesis is already limited by the upward NH$_3$ flux owing to fast chemical conversion.

Figure \ref{fig:NH3_mbar} summarizes NH$_3$ abundances at $P=1$ mbar and $100~{\rm mbar}$, which mimics the pressure levels probed by transmission and emission spectra, respectively, for various planetary masses, age, and $K_{\rm zz}$.
Overall, NH$_3$ abundances are nearly invariant with $T_{\rm eq}$ at temperate to warm exoplanets, which confirms the finding of \citet{Fortney+20}. 
Beyond a certain equilibrium temperature, photodissociation depletes NH$_3$ from the pressure levels of interest.
The threshold equilibrium temperature depends on the eddy diffusion coefficient: $T_{\rm eq}\sim 1200~{\rm K}$ for $K_{\rm zz}={10}^{11}~{\rm {cm}^2~s^{-1}}$, $T_{\rm eq}\sim 800~{\rm K}$ for $K_{\rm zz}={10}^{8}~{\rm {cm}^2~s^{-1}}$, and  $T_{\rm eq}\sim 400~{\rm K}$ for $K_{\rm zz}={10}^{5}~{\rm {cm}^2~s^{-1}}$.
Since NH$_3$ found at pressure levels higher than $100~{\rm mbar}$ is more stable to photodissociation, emission spectroscopy would have an advantage to observe quenched NH$_3$ that is less affected by photodissociation.

\subsection{Effects of different N/O, C/O, and metallicity}\label{sec:comp_effect}
We have assumed solar elemental ratios of the atmosphere thus far; however, exoplanets potentially have considerable diversity in atmospheric elemental ratios.
For example, the atmospheric N/O ratio can have both super-stellar and sub-stellar N/O, depending on the formation location and whether disk solids or gas dominates the atmospheric composition (e.g., \citealt{Piso+16,Cridland+20,Ohno&Ueda21,Turrini+22,Notsu+22}; \citetalias{Ohno&Fortney22a}).
Recent retrieval studies have reported sub-solar H$_2$O abundances in hot Jupiter atmospheres from low-resolution \citep{Pinhas+19,Welbanks+19} and high-resolution spectroscopy \citep{Pelletier+21}, which may be attributed to a high atmospheric C/O ratio.
Several studies also suggested high metallicity atmospheres on sub-Neptunes, such as GJ1214b \citep[e.g.,][]{Fortney+13,Morley+15,Ohno&Okuzumi18,Gao&Benneke18,Ohno+20,Christie+22,Kempton+23,Gao+23} and GJ436b \citep{Morley+17}, and exo-Saturns, such as WASP-39b \citep{Wakeford+18,ERS+22,ERS+22_NIRISS,ERS+22_PRISM,ERS+22_G395,ERS+22_NIRCam,ERS+22_SO2}, WASP-117b \citep{Carone+21}, \rerev{and HD149026 b \citep{Bean+23}}.

We here investigate how these different atmospheric compositions affect the NH$_3$ and HCN vertical distributions.
We test different N/O, C/O, and bulk atmospheric metallicities.
We re-compute \emph{P--T} profiles for different C/O and atmospheric metallicities, as such changes lead to important alternations in the \emph{P--T} profile.  However, for different N/O ratios we adopt the same \emph{P--T} profiles as used in the previous section, as a different N/O barely affects the abundances of other O and C bearing species, such as H$_2$O, as long as O/H and C/H are the same \citep{Hobbs+19}.
When we change N/O or C/O, we fix O/H to the solar value and change N/H or C/H to achieve the assumed N/O or C/O.
\rerev{Note that the atmospheric composition and \emph{P--T} profile can noticeably depend on individual C/H, O/H, and N/H ratios even if C/O and N/O are the same \citep{Drummond+19}.
We have fixed O/H here because we anticipate that it would be relatively easy to constrain O/H from observations, as the abundance of H$_2$O, a main atmospheric opacity source, is approximately proportional to O/H.
}
We assume a Jupiter-mass planet at the age of $1~{\rm Gyr}$ for simulations of different N/O and C/O ratios, while we assume a lower planet mass of $0.11~{\rm M_{\rm J}}$ at the age of $1~{\rm Gyr}$ for simulations of high atmospheric metallicity.

The quenched NH$_3$ abundance is sensitive to, but is not strictly proportional, to the atmospheric N/O (N/H in other words) ratio.
The top two panels of Figure \ref{fig:VULCAN_result2} show the vertical distributions of NH$_3$ and HCN abundances for N/O$=0.1\times$ and $10\times$ solar values.
The quenched NH$_3$ abundance is approximately ${10}^{-5}$ and $2\times{10}^{-4}$ for N/O$=0.1\times$ and $10\times$ solar values, respectively, for equilibrium temperature $\la 800~{\rm K}$.
Recalling that the quenched NH$_3$ abundance is $\sim 5\times{10}^{-5}$ for the solar value of N/O (see the left top panel of Figure \ref{fig:VULCAN_result}), the NH$_3$ abundance only varies by an order of magnitude across the two decades of N/O ratio.
This relatively weak dependence on N/O ratio is due to N$_2$ dominating over NH$_3$ in the deep atmospheres for a Jupiter mass planet at $1~{\rm Gyr}$, as suggested by the discrepancy between NH$_3$ and bulk N abundances in Figures \ref{fig:VULCAN_result} and \ref{fig:VULCAN_result2}.
Since this situation corresponds to $\mathcal{K}\ll 1$, Equation \eqref{eq:NH3_analytic} indicates $f_{\rm NH_3}\propto f_{\rm N}\sqrt{\mathcal{K}}\propto f_{\rm N}^{1/2}$, which explains the dependence seen in Figure \ref{fig:VULCAN_result2}.
Thus, when N$_2$ likely dominates over NH$_3$ in the deep atmosphere, one would need precise measurements of NH$_3$ to well constrain the bulk nitrogen abundance. 
The relative importance of NH$_3$ and N$_2$ in the deep atmospheres could be inferred from the second term of Equation \eqref{eq:N_diagnostic} by inserting a retrieved NH$_3$ abundance into $f_{\rm NH_3}$.

The higher C/O ratio only has minor effects on nitrogen chemistry.
The left bottom panel of Figure \ref{fig:VULCAN_result2} shows the results for C/O$=1.1$.
The NH$_3$ profile is largely similar to that for solar C/O ratio (left top panel of Figure \ref{fig:VULCAN_result}).
This is because an equilibrium NH$_3$ abundance and thus the quenched NH$_3$ abundance are insensitive to C/O \citep[see Figure 6 of][]{Moses+13b}.
The HCN profile in the upper atmosphere also becomes similar to that for solar C/O, as the HCN is initiated from the NH$_3$ photodissociation for the temperature range of our interest.
However, a higher C/O acts to increase the HCN abundance at middle to deep hot atmospheres, as the thermochemistry leads to an increased HCN abundance for high C/O. 

\rerev{Higher atmospheric metallicity leads to more N$_2$ dominant atmospheres owing to two combined effects: preference of N$_2$ as compared to NH$_3$ in thormochemical equilibrium and hot deep interior due to enhanced atmospheric opacity.}
The right bottom panel of Figure \ref{fig:VULCAN_result2} shows the results for $100 \times$ solar metallicity. 
The quenched NH$_3$ abundance is $\sim 2\times{10}^{-4}$, which is approximately two orders of magnitude lower than the actual bulk nitrogen abundance.
This large discrepancy is due to the fact that the high atmospheric metallicity yields hotter deep interiors.
We show the \emph{P--T} profiles of solar composition and 100 solar metallicity atmospheres in Figure \ref{fig:100solar}.
The figure demonstrates that the atmospheres with 100$\times$ solar metallicity yield deep adiabatic profiles much hotter than those of solar composition atmospheres, which  acts to increase $\rm N_2/NH_3$ ratio in the deep atmosphere.
Furthermore, N$_2$, a ``metal-metal'' species, is also strongly favored compared to NH$_3$ as the metallicity increases, at a given $P$ and $T$ \citep{Lodders&Fegley02,Moses+13b}. Therefore, the quenched NH$_3$ at $P\sim{1}$--$100~{\rm bar}$ starts to be depleted through thermochemical conversion at a threshold temperature of $T_{\rm eq}\sim 800~{\rm K}$, which is cooler than that found in solar composition atmospheres ($T_{\rm eq}\sim 1200~{\rm K}$).


High atmospheric metallicity also impacts the HCN abundances at the upper atmosphere.
While the HCN abundances at upper atmospheres are nearly the same as the quenched NH$_3$ abundances for solar composition atmospheres (Figure \ref{fig:VULCAN_result}), for 100$\times$ solar metallicity the HCN abundances are much lower than the quenched NH$_3$ abundances at $T_{\rm eq}\ga 600~{\rm K}$.
This finding owes to the depletion of CH$_4$ in the high metallicity atmospheres.  
The bottom panel of Figure \ref{fig:100solar} shows the vertical distributions of CH$_4$ for solar composition and 100$\times$ solar metallicity atmospheres.
We can see that\rerev{, at $T_{\rm eq}\ga 600~{\rm K}$,} the high metallicity atmosphere yields \rerev{much lower} CH$_4$ abundances \rerev{than those for solar composition atmospheres} at $P\sim 10^{-3}$--$10^{-4}~{\rm bar}$, where the HCN production takes place.
The CH$_4$ depletion is due to preference for CO over CH$_4$ at higher metallicity, as well as the hotter temperature of the higher metallicity atmospheres.
\rerev{Because HCN production requires the presence of CH$_4$ as discussed earlier, the CH$_4$ depletion results in the HCN depletion at high metallicity atmospheres}.

\section{Implications for Spectroscopic Observations}\label{sec:observation}
\subsection{\rev{Transmission spectroscopy}}
\begin{figure*}[t]
\centering
\includegraphics[clip, width=0.5\hsize]{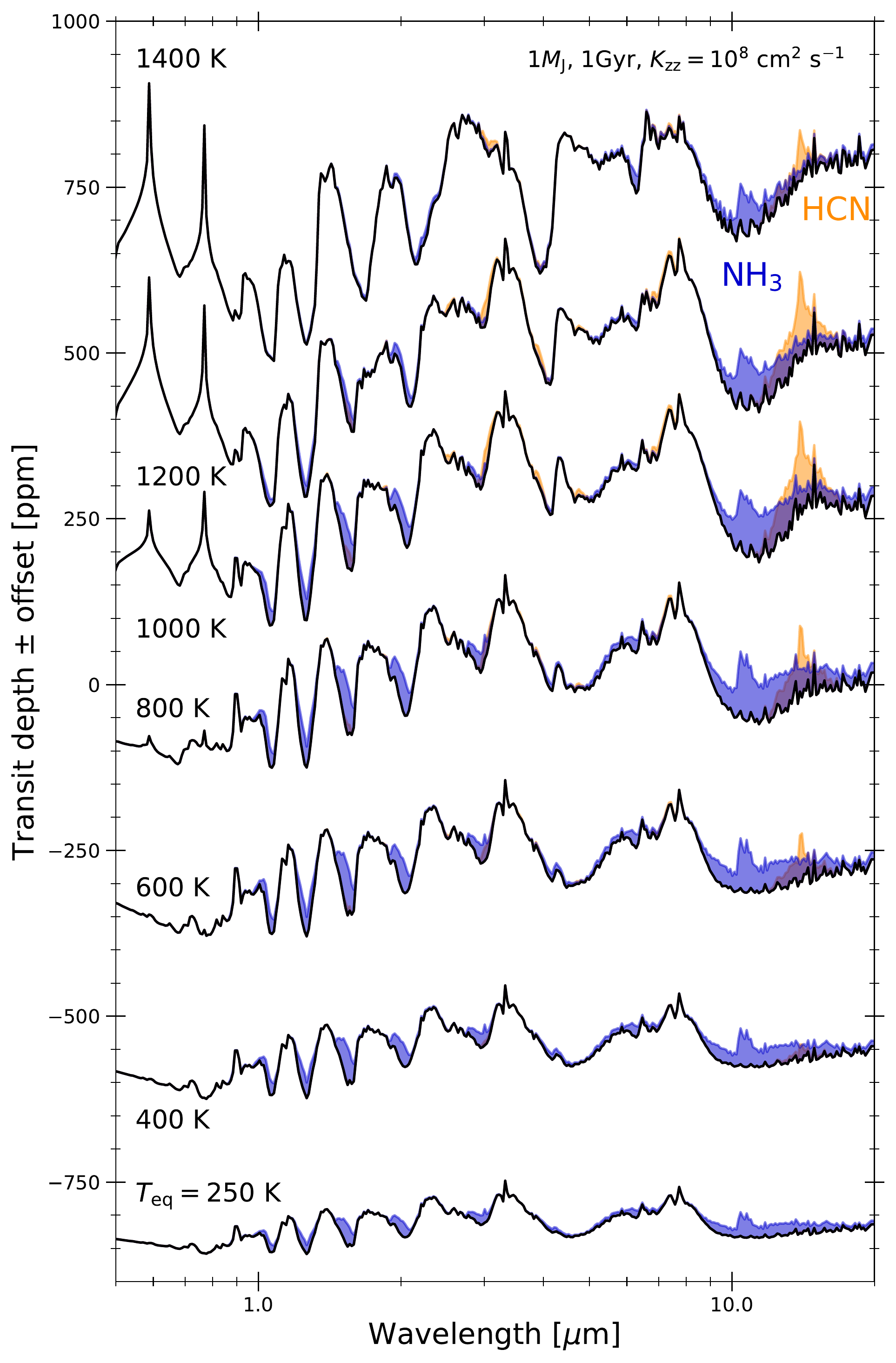}
\includegraphics[clip, width=0.49\hsize]{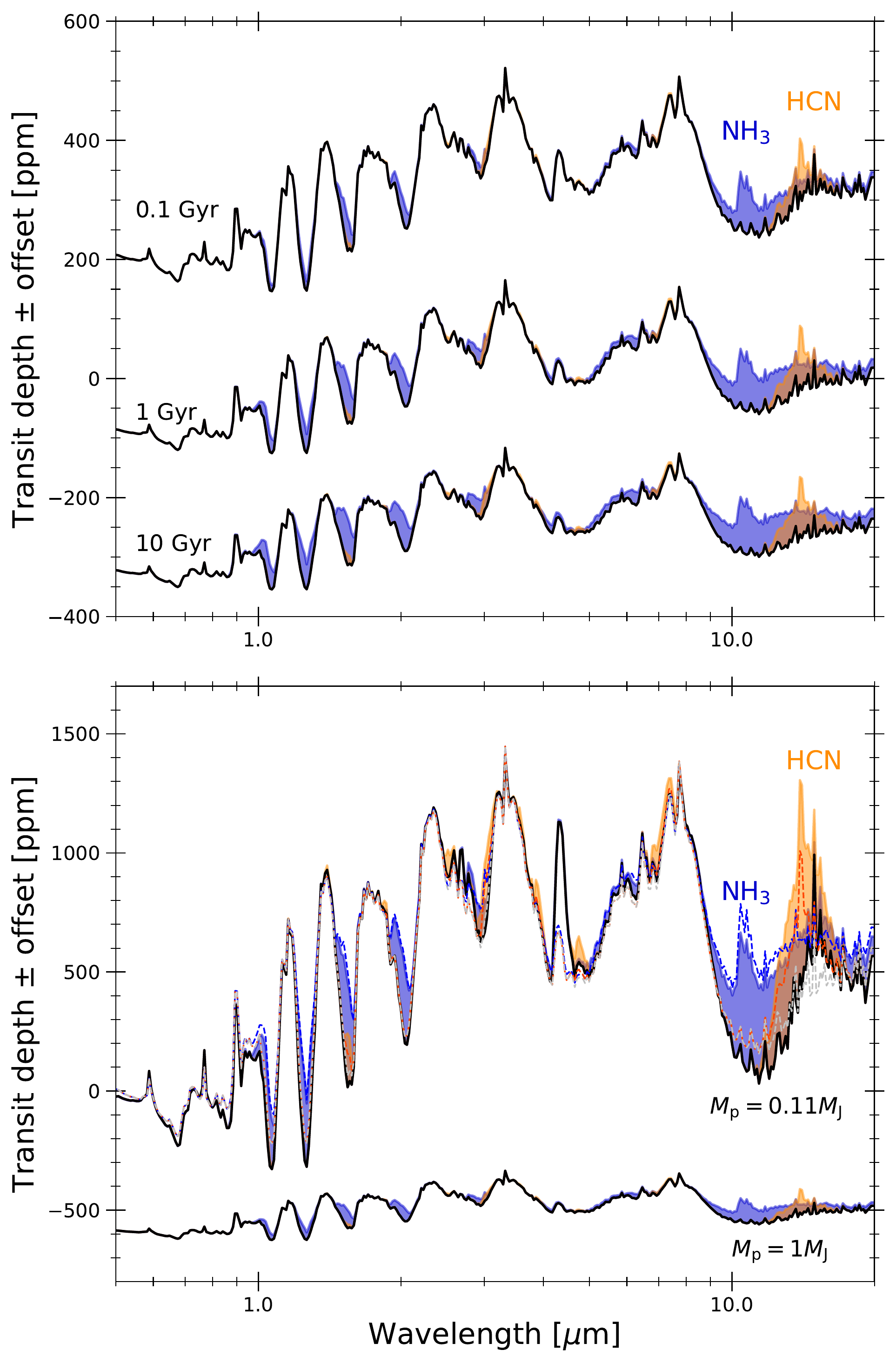}
\caption{Synthetic transmission spectra of warm exoplanets. The left panel shows the transit depths of Jupiter-mass planets orbiting around a Sun-like star with various equilibrium temperature, where we assume $1~{\rm Gyr}$ age. The black lines show the spectra without NH$_3$ and HCN, and the blue and orange shaded parts denote the excess transit depth when we include NH$_3$ and HCN absorption.
\rev{The upper right panel shows the spectra of Jupiter-mass planets with $T_{\rm eq}=800~{\rm K}$ at different system ages. The lower right panel shows the spectra for sub-Jupiter ($M_{\rm p}=0.11~{\rm M_{\rm J}}$) and Jupiter mass ($1~{\rm M_{\rm J}}$) planets with $T_{\rm eq}=800~{\rm K}$ at $1~{\rm Gyr}$ age. We assume solar composition atmospheres and $K_{\rm zz}={10}^{8}~{\rm {cm}^2~s^{-1}}$ for all spectra shown here.}
}
\label{fig:trans}
\end{figure*}
\begin{figure*}[t]
\centering
\includegraphics[clip, width=\hsize]{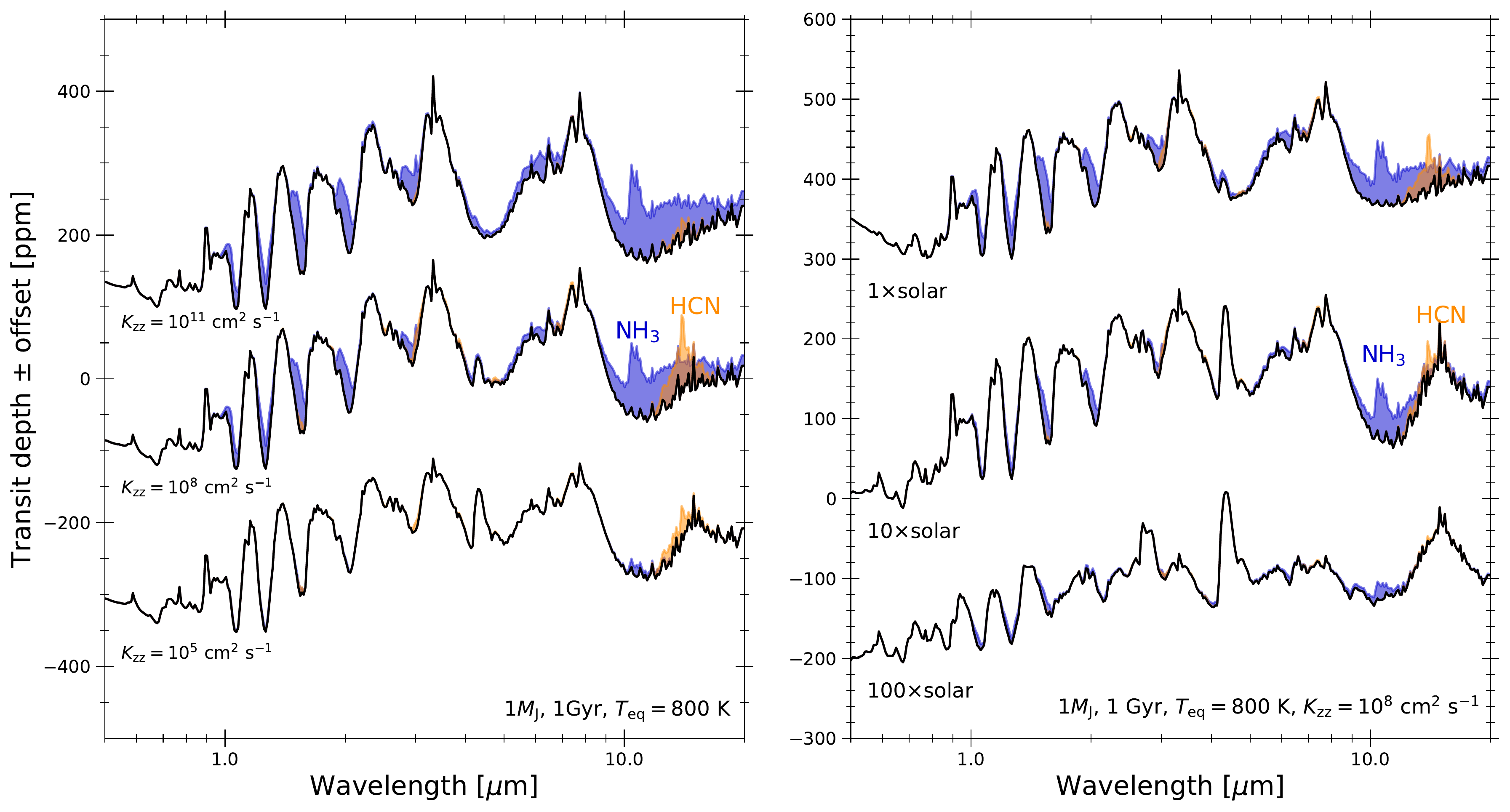}
\caption{\rev{Transmission spectra of a 1 Gyr old Jupiter-mass planet with $T_{\rm eq}=800~{\rm K}$ around Sun-like stars for various $K_{\rm zz}$ values (left) and atmospheric metallicities (right).}
}
\label{fig:trans2}
\end{figure*}

We next turn to the study of the observational feasibility of NH$_3$ and HCN in exoplanet transmission spectroscopy.
We use the publicly available code CHIMERA \citep{Line+13} to generate synthetic transmission spectra. 
We customize CHIMERA so that the molecular abundances of H$_2$O, CH$_4$, CO, CO$_2$, NH$_3$, N$_2$, HCN, C$_2$H$_2$, C$_2$H$_6$ are extracted from the results of VULCAN.
We assume thermochemical equilibrium for the remaining chemical species.
We assume Jupiter-mass planets with 10 bar reference radii of $1.21$, $1.11$, and $1.04 R_{\rm J}$ at $0.1$, $1$, and $10~{\rm Gyr}$, respectively.
These radii correspond to the surface gravities of $16.90$, $19,96$, and $22.71~{\rm m~s^{-2}}$, assumed in our photochemical calculations (see Table \ref{table:1}).
To compute transit depths, we assume a solar radius for all spectral calculations.

The magnitudes of NH$_3$ and HCN features depend on the planetary equilibrium temperatures.
The left top panel of Figure \ref{fig:trans} shows the transmission spectra for different $T_{\rm eq}$ values.
In general, NH$_3$ leaves relatively strong spectral features at $1.2$, $1.5$, $2.0$, $3.0$, and $11~{\rm {\mu}m}$, while HCN leaves features only at $14~{\rm {\mu}m}$.
The NH$_3$ features at near-infrared wavelength are in agreement with \citet{MacDonald&Madhusudhan17}, whereas several unique features beyond $2~{\rm {\mu}m}$ suggested by \citet{MacDonald&Madhusudhan17} are barely seen in our spectra.
This is because we mainly focus on warm exoplanets in which the spectral features of CH$_4$ tend to obscure the NH$_3$ spectral features.
In our spectra relevant to hot Jupiters ($T_{\rm eq}=1200$ and $1400~{\rm K}$), the NH$_3$ depletion through thermochemical conversion and photodissociation significantly weakens NH$_3$ feature, as compared to \citet{MacDonald&Madhusudhan17} who fixed the NH$_3$ abundance regardless of planetary equilibrium temperature.

We also conduct the same analysis for HCN; however, HCN features are almost absent in our synthetic spectra.
HCN features are weak in our models because HCN is present only at low pressures ($\la {10}^{-3}$~{\rm bar}) as HCN forms through photodissociation of NH$_3$ at $P\sim {10}^{-3}$--${10}^{-4}$~{\rm bar}.
The magnitudes of HCN features at $3.1$ and $14~{\rm {\mu}m}$, though weak, gradually increase with increasing equilibrium temperature, as the enhanced stellar UV photons cause the HCN production at slightly higher pressures. 
In our model, the $3.1~{\rm {\mu}m}$ feature is at most ${\sim}20~{\rm ppm}$, while the $14~{\rm {\mu}m}$ feature could be larger than $50~{\rm ppm}$.

The NH$_3$ and HCN features also modestly depend on the planet's age.
The right top panel of Figure \ref{fig:trans} shows the transmission spectra of Jupiter-mass planets with $T_{\rm eq}=800~{\rm K}$ and $K_{\rm zz}={10}^{8}~{\rm {cm}^2~s^{-1}}$ for different system ages.
As discussed in Section \ref{sec:result_age}, NH$_3$ and HCN abundances are lower at younger planets because of hotter interiors and deep atmospheres \citep{Fortney+20}.
Thus, the NH$_3$ and HCN features tend to be weaker for younger planets.
On the other hand, younger planets have lower surface gravity and larger atmospheric scale height.
Because the large scale height enhances the amplitudes of spectral features and partly compensates the effect of decreased abundances of NH$_3$ and HCN, the magnitudes of spectral features only moderately depend on system age.

\rev{
We note that, of course, the actual strength of NH$_3$ and HCN features depend on the planetary gravity.
The right bottom panel of Figure \ref{fig:trans} shows the spectra for the planetary mass of $1M_{\rm J}$ and $0.11M_{\rm J}$, corresponding to the surface gravity of $19.96$ and $3.47~{\rm m~s^{-2}}$ at 1 Gyr age (see Table \ref{table:1}). 
The $0.11M_{\rm J}$ mass planet shows much larger spectral features, including NH$_3$ and HCN features.
This is mostly because the spectral features of transmission spectrum are scaled by atmospheric scale height \citep[e.g.,][]{Kreidberg18}.
In fact, the two spectra agree well with each other once we rescale the $1M_{\rm J}$ planet spectrum by multiplying a factor of $19.969/3.475$, the ratio of the surface gravities (dashed lines).
The different gravities still alter the spectral shapes at certain wavelength owing to differences of atmospheric composition, such as the  CO$_2$ absorption feature at $4.3~{\rm{\mu}m}$.
}

The magnitudes of NH$_3$ and HCN features also depend on the eddy diffusion coefficient.
The left panel of Figure \ref{fig:trans2} shows the transmission spectra for different $K_{\rm zz}$ values.
The NH$_3$ features become prominent as $K_{\rm zz}$ increases, as a higher $K_{\rm zz}$ pushes the depth of the photodissociation to lower pressures.
On the other hand, the HCN features have a non-monotonic dependence on $K_{\rm zz}$.
The HCN features become weak at high $K_{\rm zz}$ because HCN production, which is driven by NH$_3$ photodissociation, takes place at much higher altitudes than that probed by transmission spectrum.
The features also become weak at very low $K_{\rm zz}$, as thermochemistry tends to convert NH$_3$ to N$_2$ in the middle atmospheres before NH$_3$ is converted to HCN.
Thus, interestingly, there is a sweet spot value of $K_{\rm zz}$ that maximizes the HCN features.

\rev{
Atmospheric metallicity also has a strong impact on the NH$_3$ and HCN feature strengths.
The right panel shows the spectra for the atmospheric metallicities of $1\times$, $10\times$, and $100\times$ the solar value.
The NH$_3$ and HCN features become weaker as the metallicity becomes higher. 
This is because the NH$_3$ abundance remains roughly the same as  metallicity increases owing to the conversion of N into N$_2$ at deep atmospheres (see \citetalias{Ohno&Fortney22a}, Section \ref{sec:N_map}, and Section \ref{sec:comp_effect}) whereas other molecular abundances, such as H$_2$O, increase with increasing metallicity.
As a result, NH$_3$ features tend to be embedded into the absorption features of other molecules.
Meanwhile, \rerev{unless the planet is as cool as $T_{\rm eq}\la 400~{\rm K}$}, higher atmospheric metallicity leads a lower abundance of CH$_4$, which results in the depletion of HCN (Section \ref{sec:comp_effect},  see also Figure \ref{fig:100solar}). Thus, observations of NH$_3$ and HCN are more challenging for higher metallicity atmospheres.
}

\subsection{\rev{Emission spectroscopy}}
\begin{figure*}[t]
\centering
\includegraphics[clip, width=0.5\hsize]{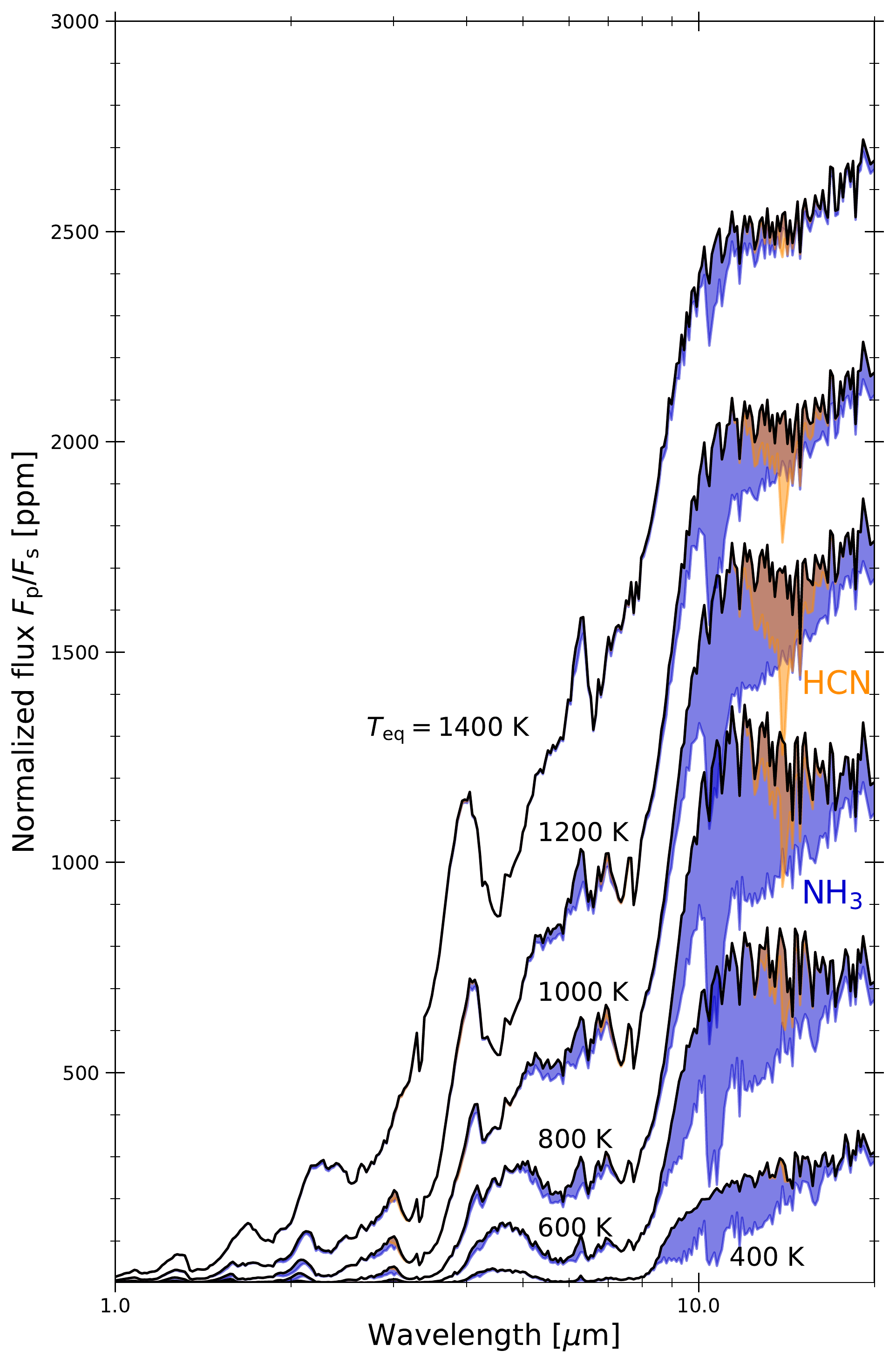}
\includegraphics[clip, width=0.49\hsize]{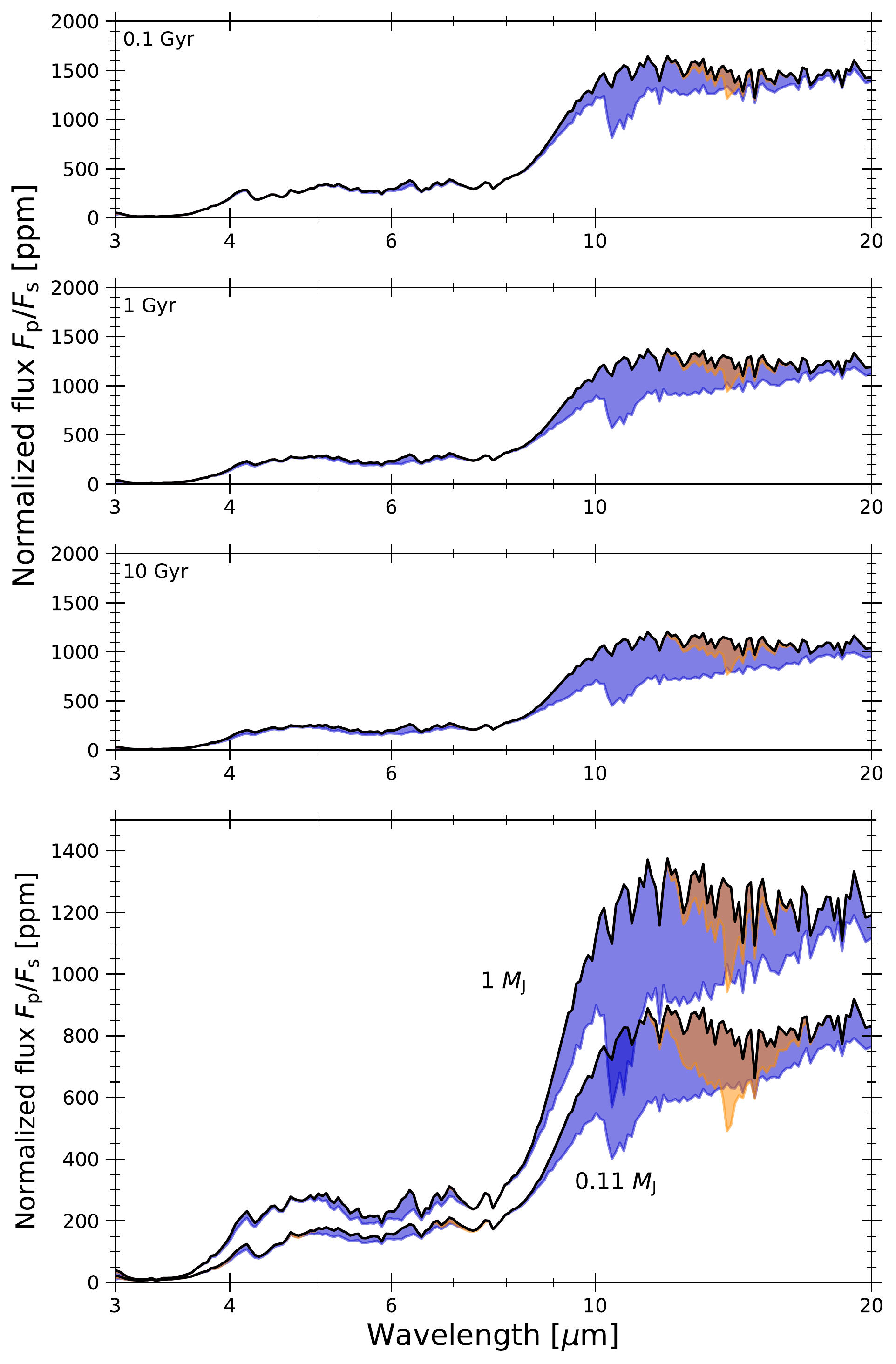}
\caption{\rev{Synthetic emission spectra of giant planets around a Sun-like star. The black lines show the spectra without NH$_3$ and HCN, and the blue and orange shaded parts denote the reduction of planetary thermal emission when we include NH$_3$ and HCN absorption. The left panel shows the spectra of Jupiter-mass planets at 1 Gyr age for various planetary equilibrium temperature. The right top three panels show the spectra of Jupiter-mass planets with $T_{\rm eq}=800~{\rm K}$ at different ages. The right bottom panel shows the spectra for $T_{\rm eq}=800~{\rm K}$ and 1 Gyr age at different planetary masses. We have assumed solar composition atmospheres and $K_{\rm zz}={10}^{8}~{\rm cm^2~s^{-1}}$ for all spectra presented here.} 
}
\label{fig:emis}
\end{figure*}
\begin{figure*}[t]
\centering
\includegraphics[clip, width=\hsize]{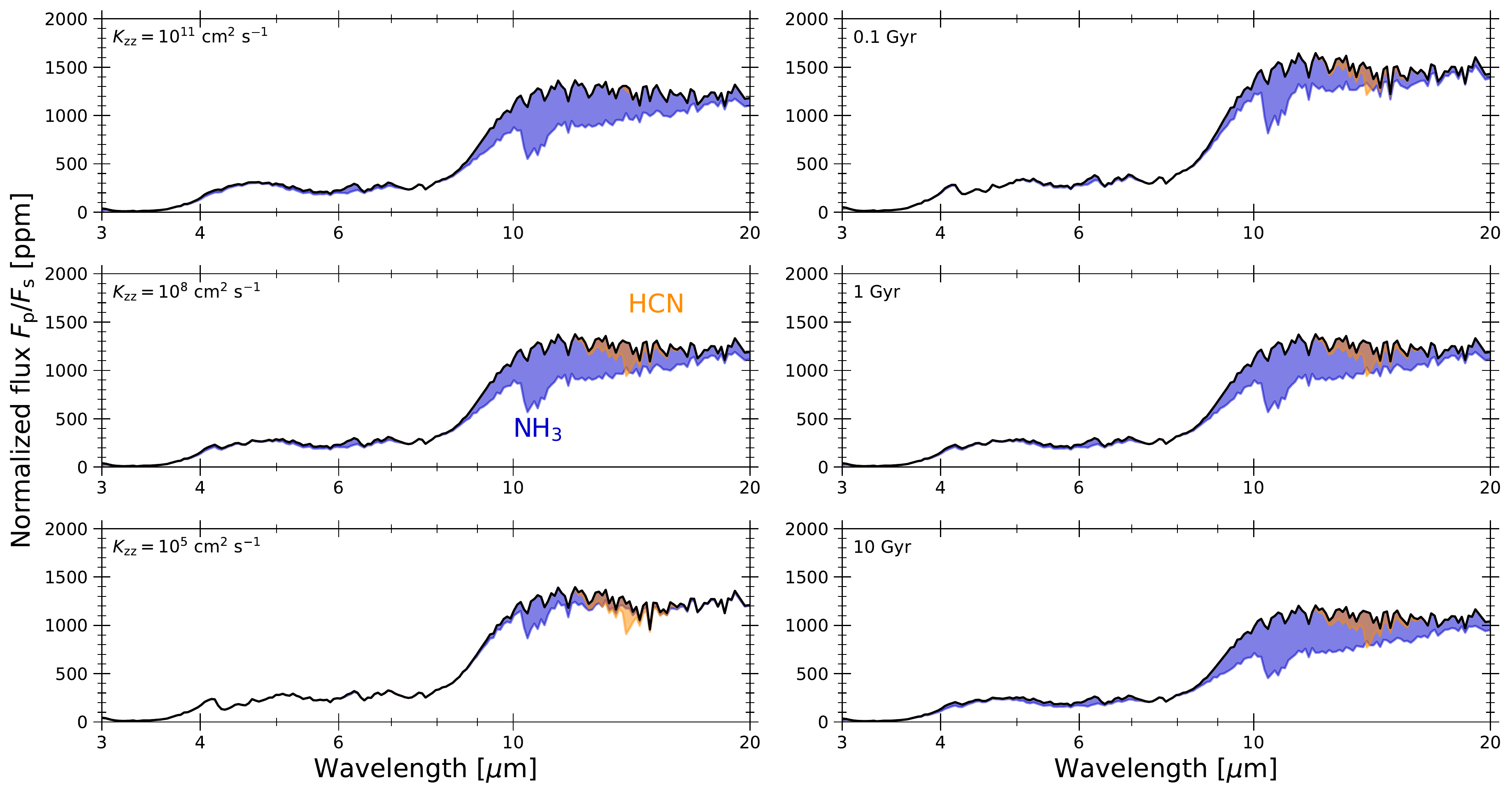}
\caption{\rev{Emission spectra of 1 Gyr old Jupiter-mass planets with $T_{\rm eq}=800~{\rm K}$ around Sun-like stars for various values of eddy diffusion strength $K_{\rm zz}$ (left column) and atmospheric metallicity (right column).} 
}
\label{fig:emis2}
\end{figure*}
\rev{
We investigate the observational feasibility of NH$_3$ and HCN in emission spectroscopy as well. As in the previous section, we use the CHIMERA to compute synthetic emission spectra for hypothetical Jupiter-mass planets around sun-like star, where the stellar spectrum is taken from the PHOENIX grid \citep{Husser+13} for stellar effective temperature of $T_{\rm eff}=5780~{\rm K}$, metallicity of $[M/H]=0$, and gravity of $\log{(g)}=4.43$. 
}

\rev{
The NH$_3$ features in the emission spectra are maximized at warm exoplanets with $T_{\rm eq}\sim600$--$1000~{\rm K}$, similar to transmission spectroscopy.
Figure \ref{fig:emis} shows the emission spectra for various equilibrium temperatures.
The emission spectra always exhibit the strongest NH$_3$ feature around $11~{\rm \mu m}$ followed by the feature around $\sim6~{\rm {\mu}m}$. 
The features at longer wavelengths are more noticeable because of the more favorable planet-to-star flux contrast.
Since the planetary emitting flux is intrinsically weaker at cooler planets, as seen in the $T_{\rm eq}=400~{\rm K}$ case, the cooler planets have weaker NH$_3$ features if we assume the same stellar properties.
On the other hands, hotter planets emit more flux but have a lower NH$_3$ abundance owing to thermal conversion, which also results in weaker NH$_3$ features as seen in the cases of $T_{\rm eq}=1200$ and $1400~{\rm K}$.
}

\rev{
The system age moderately affects NH$_3$ and HCN features in emission spectra.
The upper three right panels of Figure \ref{fig:emis} show the spectrum of a Jupiter-mass planet with $T_{\rm eq}=800~{\rm K}$ at $0.1$, $1$, and $10$ Gyr.
The planet at $0.1$ Gyr has relatively weak NH$_3$ feature because their hot deep atmospheres convert NH$_3$ into N$_2$ efficiently.
On the other hand, the planetary emission itself becomes stronger at younger ages owing to the hot deep atmosphere.
}

\rev{
The strength of NH$_3$ feature also depends on the eddy diffusion coefficient. 
The left columns of Figure \ref{fig:emis2} show how the strength of NH$_3$ and HCN features vary with $K_{\rm zz}$ for a 1 Gyr old Jupiter-mass planet with $T_{\rm eq}=800~{\rm K}$.
The spectral shape is nearly the same between $K_{\rm zz}={10}^{11}$ and ${10}^{8}~{\rm {cm}^2~s^{-1}}$.
This is because the quenched NH$_3$ abundance is insensitive to $K_{\rm zz}$ (\citealt{Saumon+06,Zahnle+14,Fortney+20}; \citetalias{Ohno&Fortney22a}), and emission spectra probe the pressure level below the NH$_3$ photodissociation base at $K_{\rm zz}={10}^{8}~{\rm {cm}^2~s^{-1}}$.
Thus, the further elevation of the photodissociation base by high $K_{\rm zz}$ no longer affects the strength of NH$_3$ feature.
NH$_3$ features become weaker at very low value of $K_{\rm zz}={10}^{8}~{\rm {cm}^2~s^{-1}}$ because thermochemistry converts NH$_3$ to N$_2$ at the pressure level probed by emission spectra.
}

\rev{
Atmospheric metallicity substantially affects the strength of NH$_3$ and HCN features.
The right columns of Figure \ref{fig:emis2} show the emission spectra at $1\times$, $10\times$, and $100\times$ solar metallicity.
The NH$_3$ and HCN feature becomes weaker at the higher atmospheric metallicity owing to the thermochemical conversion of NH$_3$ into N$_2$ at deep atmospheres, which results in the depletion of NH$_3$ and HCN at upper atmospheres.
This trend is the same as that seen in the transmission spectroscopy.
However, while the high metallicity weakens the entire spectral features for transmission spectrum, the high metallicity instead enhance planetary emission by warming the atmosphere.
}

\rev{
As in the transmission spectra, HCN features are relatively weak compared to NH$_3$ features. 
Only the feature around $14~{\rm {\mu}m}$ is noticeable.
The relatively weak HCN feature can be attributed to the vertical distribution of HCN.
The emission spectrum probes deeper in the atmosphere than the transmission spectrum, while HCN is mainly produced by photochemistry in the upper atmosphere.
This spatial separation between the HCN production region and observationally sensitive region results in the relatively weak HCN features in the emission spectrum.
}

\subsection{\rev{Observational Feasibility of NH$_3$ and HCN}}\label{sec:feasibility}
\begin{figure*}[t]
\centering
\includegraphics[clip, width=\hsize]{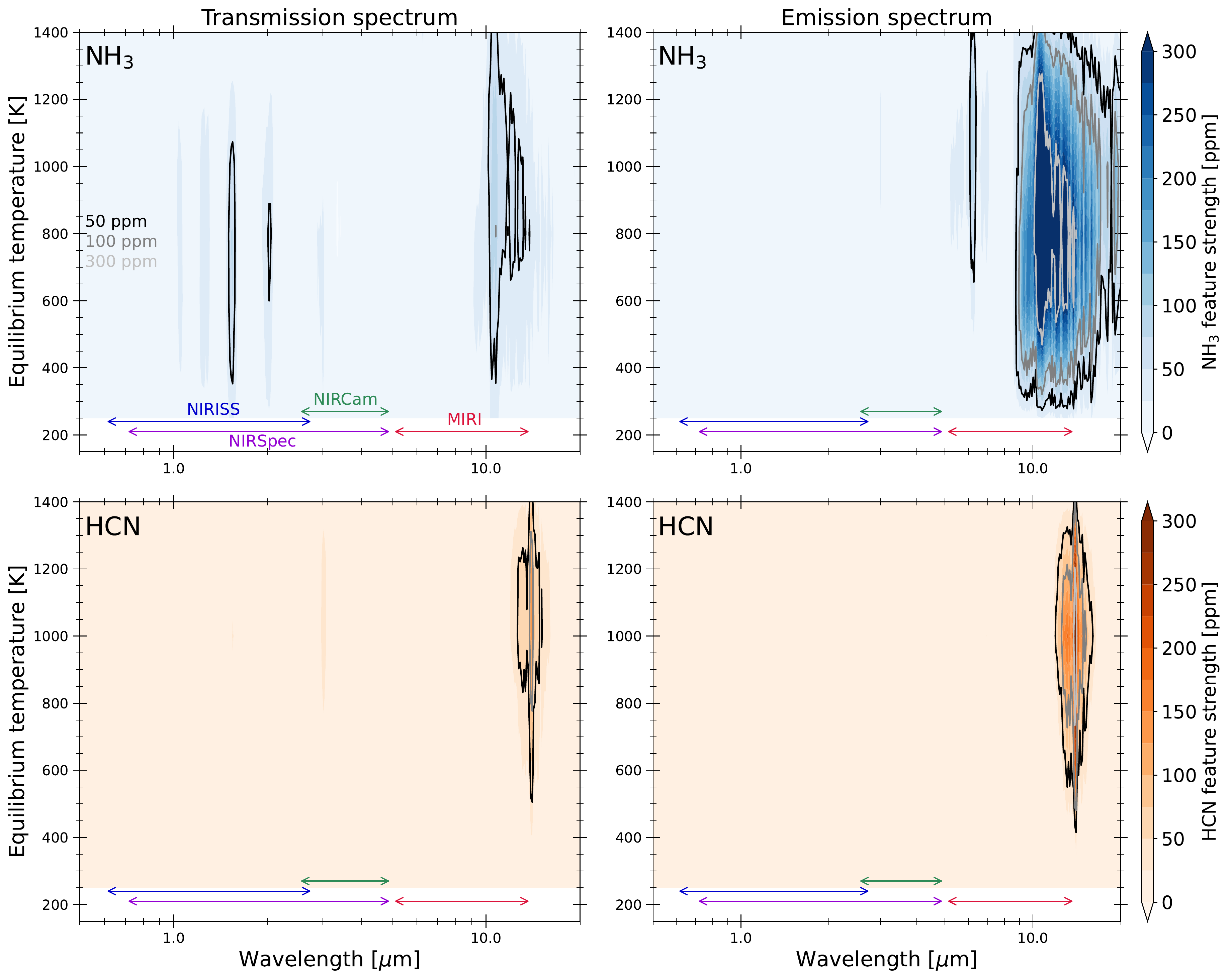}
\caption{\rev{Strength of NH$_3$ (top row) and HCN (bottom row) absorption features as a function of wavelength and planetary equilibrium temperature (see Section \ref{sec:feasibility} for definition of the feature strength). The left and right columns show the results for transmission and emission spectra, respectively. The black, gray, and silver lines denote the contours of 50, 100, and 300 ppm, respectively. The colored arrows at the bottom of each panel denote the wavelength range of JWST NIRISS (0.6--2.8 $\rm {\mu}m$), NIRCam (2.5--5 $\rm {\mu}m$), NIRSpec (0.7--5 $\rm {\mu}m$), and MIRI (5--14 $\rm {\mu}m$), taken from \citet{Batalha+17}. We have assumed a Jupiter-mass planet around a Sun-like star at 1 Gyr age with a solar composition atmosphere and $K_{\rm zz}={10}^{8}~{\rm {cm}^2~s^{-1}}$. }
}
\label{fig:metric}
\end{figure*}

We quantify the observational feasibility of NH$_3$ by examining the difference in spectrum with NH$_3$ and without nitrogen species (i.e., NH$_3$ and HCN).
\rev{For transmission spectra, we define the strength of NH$_3$ features as}
\begin{equation}
    \Delta \left(\frac{R_{\rm p}}{R_{\rm s}}\right)^2_{\rm NH_3}=\left(\frac{R_{\rm p}}{R_{\rm s}}\right)^2_{\rm w/~NH_3}-\left(\frac{R_{\rm p}}{R_{\rm s}}\right)^2_{\rm w/o~NH_3\&HCN}.
\end{equation}
This approach is similar to \citet{MacDonald&Madhusudhan17}.
\rev{For emission spectra, the strength is defined as}
\begin{equation}
    \Delta \left(\frac{F_{\rm p}}{F_{\rm s}}\right)_{\rm NH_3}=\left(\frac{F_{\rm p}}{F_{\rm s}}\right)_{\rm w/o~NH_3\&HCN}-\left(\frac{F_{\rm p}}{F_{\rm s}}\right)_{\rm w/~NH_3}.
\end{equation}
\rev{We compute the above metric at wavelength for various planetary equilibrium temperatures assuming a Jupiter-mass planets around a Sun-like star at 1 Gyr age, with a solar composition atmosphere and $K_{\rm zz}={10}^{8}~{\rm cm^2~s^{-1}}$.
We note that the actual feature strength depends on stellar and planetary properties; for example, the feature strengths of the transmission spectrum are enhanced for smaller stars and/or lower gravity planets with larger scale heights.
Our objective here is to explore the optimum range of wavelength and planetary equilibrium temperature for detecting NH$_3$ under the same system conditions.
}
The same analysis is also carried out for HCN.

The left column of Figure \ref{fig:metric} shows the NH$_3$ and HCN feature strengths \rev{of the transmission spectrum} as a function of wavelength and equilibrium temperature.
In the transmission spectra, NH$_3$ leaves relatively strong spectral features at $1.5$, $2.0$, and  $11.0~{\rm {\mu}m}$.
These features could be relatively strong, at an amplitude of $>50~{\rm ppm}$ (black contours), at equilibrium temperatures from $\sim 400$--$1000~{\rm K}$. Since the observational noise floor was anticipated to be $20$--$50$ ppm \citep{Greene+16}, the result suggests that NH$_3$ could be identified in the near infrared with JWST.
The unique NH$_3$ feature at $11.0~{\rm {\mu}m}$, accessible to JWST MIRI, would be particularly useful to identify NH$_3$, as the feature is present in an opacity window of H$_2$O and CH$_4$ and atmospheric clouds/hazes tend to be optically thin at \rerev{longer wavelengths}.
In general, warm ($\sim 400$--$1000~{\rm K}$) exoplanets are optimum targets for detecting NH$_3$ because hot planets \rerev{tend to experience} NH$_3$ depletion by both thermochemistry and photochemistry whereas cold planets suffer from weak spectral features owing to small atmospheric scale heights.

\rev{
Although the NH$_3$ feature strength of the transmission spectra is at most  $\sim 50~{\rm ppm}$ for Jupiter-mass planets around Sun-like stars, the emission spectrum could show much larger NH$_3$ features for the same conditions.
The right column of Figure \ref{fig:metric} shows the NH$_3$ and HCN feature strength for emission spectra.
In emission, NH$_3$ leaves a moderately strong feature at $\sim6~{\rm {\mu}m}$ and very prominent feature at $\sim11~{\rm {\mu}m}$.
In particular, the $11~{\rm {\mu}m}$ feature strength exceeds $300~{\rm ppm}$ at $T_{\rm eq}\sim 400$--$1200~{\rm K}$, meaning that the NH$_3$ feature is more than $6$ times larger than the noise floor of JWST-MIRI anticipated by \citet{Greene+16}. 
Although the observational noise tends to be large at such long wavelength owing to the decreased number of stellar photons, given the prominence of feature and expectation for low cloud and haze opacity, emission spectroscopy by JWST-MIRI would offer a viable window to search for NH$_3$ in wide range of the equilibrium temperatures through the $11~{\rm {\mu}m}$ feature.
}

\rev{
Our model predicts that HCN features are very weak at a wide range of wavelengths and equilibrium temperatures.
In both transmission and emission spectra, the HCN feature strength is smaller than $50~{\rm ppm}$ over the entire $T_{\rm eq}$ range explored by this study (250--1400~{\rm K}), except for the feature around $14~{\rm {\mu}m}$ whose feature strength could be as large as $50~{\rm ppm}$ for transmission and $300~{\rm ppm}$ for emission spectra.
Thus, the observation by JWST-MIRI may still be able to find a hint of HCN feature in addition to the NH$_3$ feature.
On the other hand, if future observations detect prominent HCN features at $<10~{\rm {\mu}m}$, it potentially indicates that the atmospheric composition is significantly different from the solar composition assumed in these figures.
For example, a very high C/O ratio (C/O$\gg$1) increases HCN abundance by orders of magnitude \citep{Moses+13,Hobbs+21}, which may yield noticeable HCN features at short wavelengths.
}

\section{Discussion}\label{sec:discussion}


\subsection{Effects of Photochemical Haze}\label{sec:discussion_haze}
We have assumed clear atmospheres in this study; however, recent studies suggested that warm exoplanets are universally veiled by photochemical hazes \citep[e.g.,][]{Crossfield&Kreidberg17,Gao+20,Dymont+21}.
Photochemical hazes can affect our results in multiple ways, including flattening spectral features that limits the observability of atmospheric molecules \citep[e.g.,][]{Fortney05,Morley+13,Lavvas&Koskinen17,Kawashima&Ikoma18,Kawashima&Ikoma19,Adams+19,Lavvas+19,Ohno&Kawashima20,Gao&Zhang20,Steinrueck+21,Steinrueck+23,Ohno&Tanaka21,Arfaux&Lavvas22}.

\rev{An additional intriguing effect is that} hazes alter the energy balance and \emph{P--T} profiles significantly \citep{Morley+15,Lavvas&Aufaux21,Arfaux&Lavvas22,Steinrueck+23}.
For example, \citet{Molaverdikhani+20} suggested that (though they focused on radiative feedback of mineral clouds) thick clouds may cause hot deep atmospheres and deplete NH$_3$.
\rev{The consequence of radiative feedback of photochemical haze is not trivial. 
The hazes heat upper atmospheres by absorbing stellar flux \citep{Morley+15,Lavvas&Aufaux21}, which may deplete NH$_3$ via thermochemical conversion. On the other hand, the hazes cool lower atmospheres by blocking stellar light from reaching the deeper atmosphere \citep{Morley+15}, which may act to maintain the quenched NH$_3$ abundance by suppressing the thermochemical conversion of NH$_3$ to N$_2$. 
}

Hazes also shield atmospheres from stellar UV photons and may alter photochemistry.
\citet{Sagan&Chyba97} and \citet{Wolf&Toon10} suggested that photochemical hazes could shield NH$_3$ from solar UV photons in the Archean Earth, which might help to maintain warm climates.
If similar processes work at warm exoplanets, hazes may help to stabilize NH$_3$ against photodissociation.
Overall haze effects depend on particle properties, such as particle size and shapes, which are controlled by microphysical process.
We will comprehensively investigate the overall impacts of photochemical hazes on atmospheric structure and chemistry in future work.

\subsection{Effects of Stellar Spectral Type}\label{sec:discussion_stellar}
\begin{figure}[t]
\centering
\includegraphics[clip, width=\hsize]{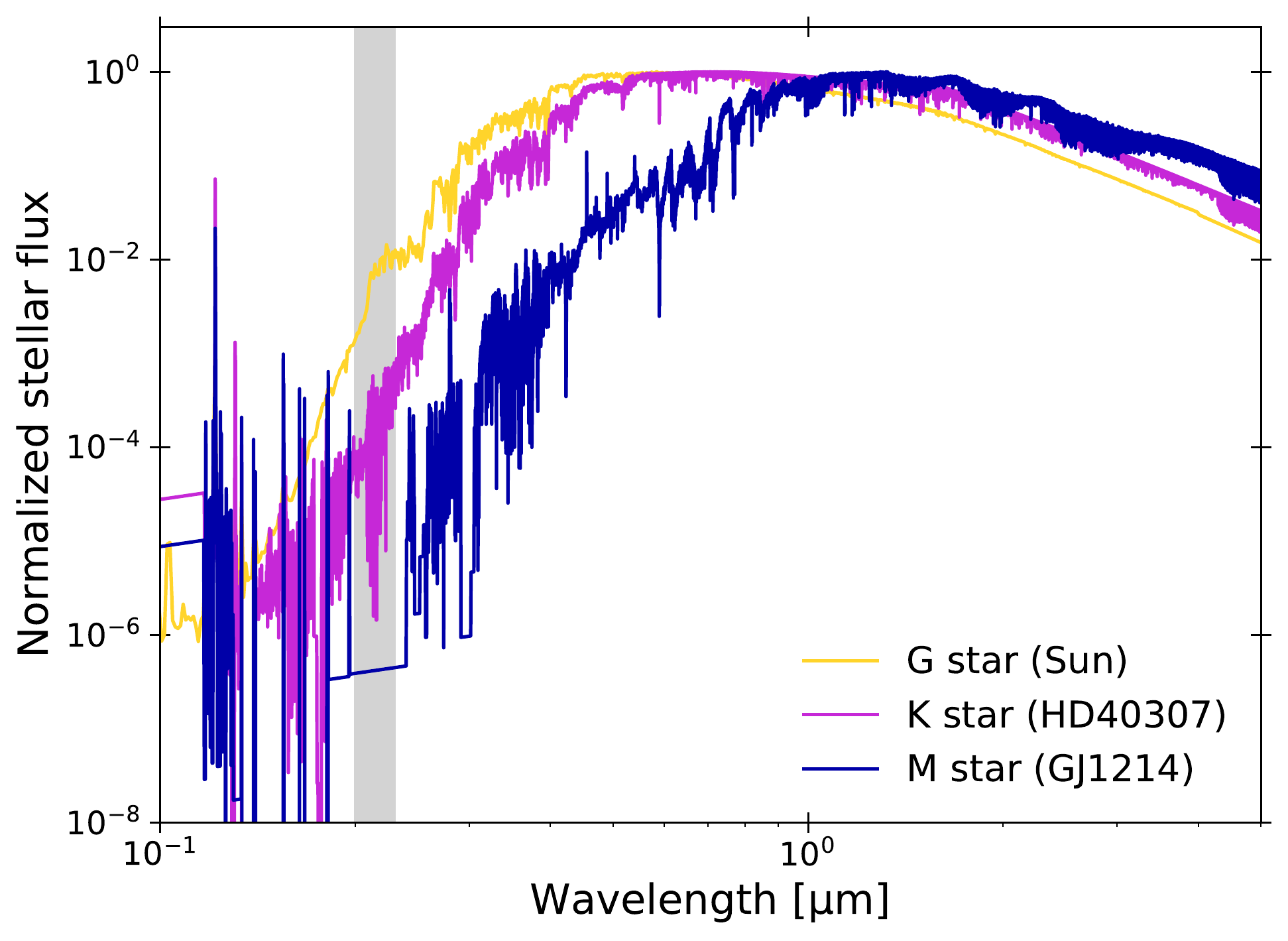}
\includegraphics[clip, width=\hsize]{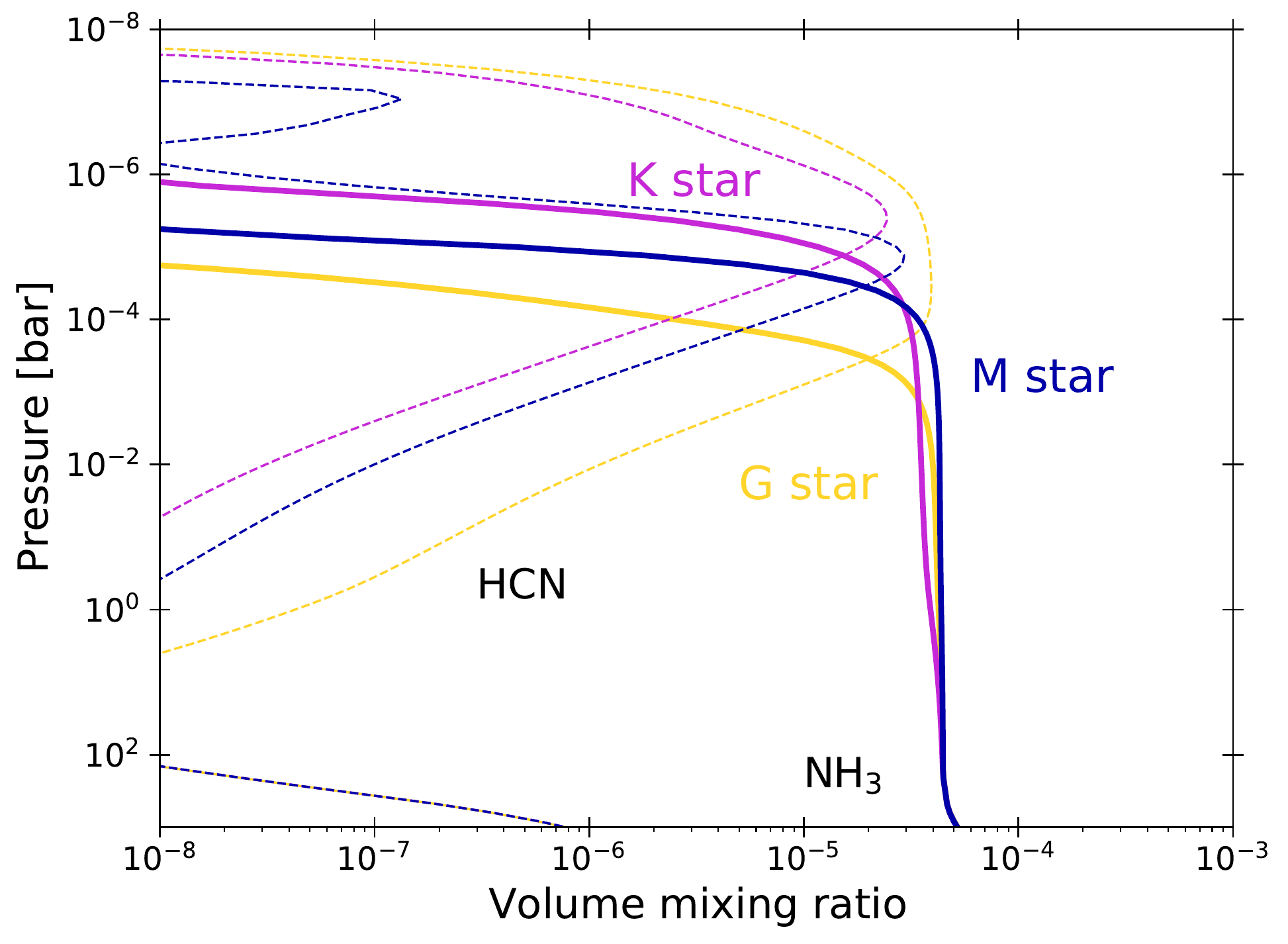}
\caption{(Top) Stellar spectra $\lambda F_{\rm \lambda}$ at the stellar surface. The yellow, magenta, and navy lines show the spectra of G-type (Sun), K-type (HD40307), and M-type (GJ1214) stars, respectively. The flux are normalized by the maximum value. The gray shaded region denotes the wavelength band of $200$--$230~{\rm nm}$, which mainly drives NH$_3$ photodissociation.
(Bottom) Vertical distributions of NH$_3$ (solid lines) and HCN (dashed lines) on warm planets around the different spectral type stars listed in the top panel. We assume a Jupiter-mass planet with age of $1~{\rm Gyr}$, $T_{\rm eq}=600~{\rm K}$, and $K_{\rm zz}={10}^{8}~{\rm {cm}^2~s^{-1}}$. 
}
\label{fig:stellar_effect}
\end{figure}
\begin{figure}[t]
\centering
\includegraphics[clip, width=\hsize]{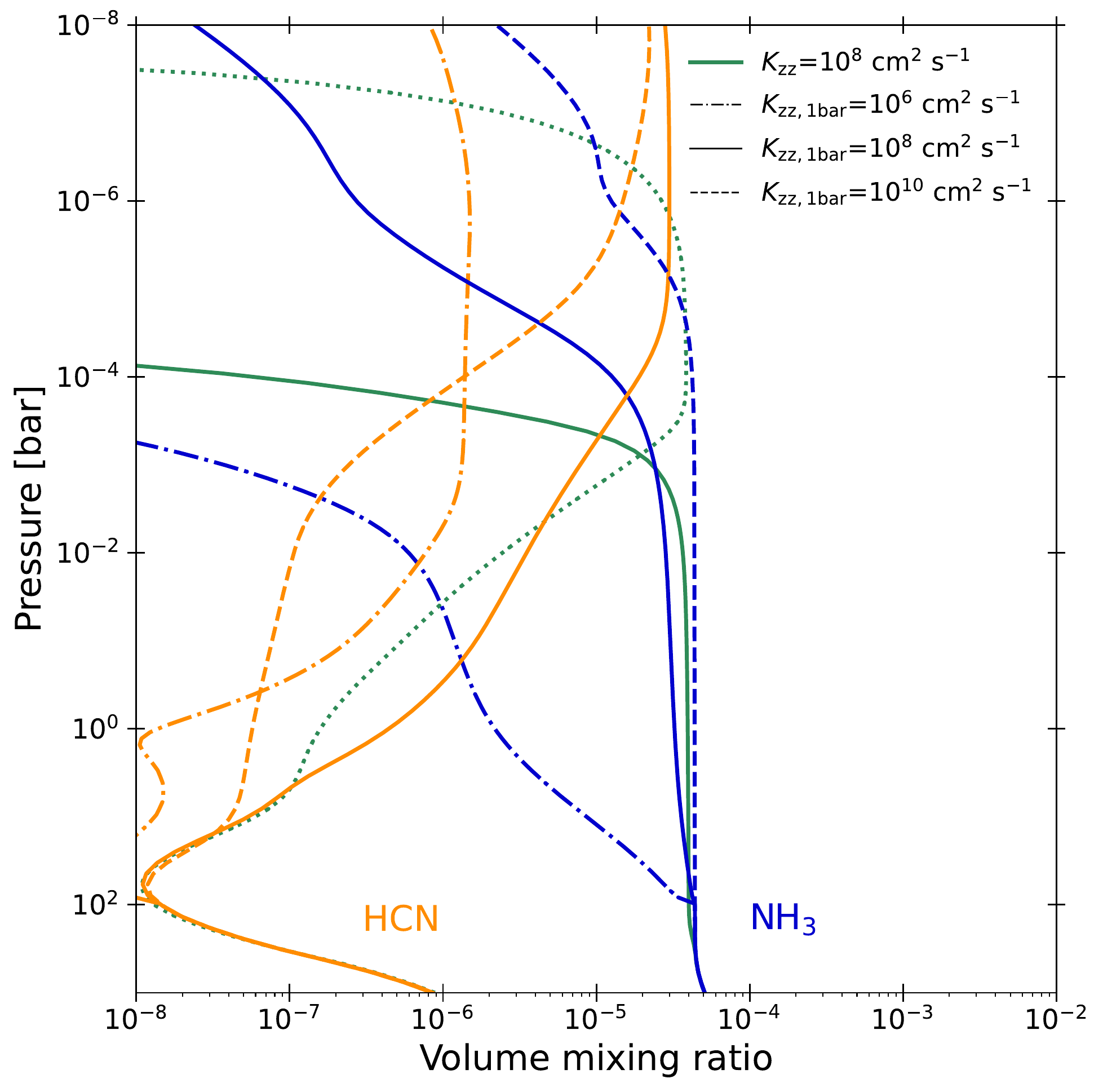}
\caption{\rev{Vertical distributions of NH$_3$ and HCN for different  parameterizations of $K_{\rm zz}$. The blue and orange dash-dot, solid, dashed lines show the vertical distribution of NH$_3$ and HCN for vertically varying $K_{\rm zz}$ with $K_{\rm zz,1~{\rm bar}}={10}^{6}$, $10^{8}$, and $10^{10}~{\rm cm^2~s^{-1}}$, respectively. The green solid and dotted lines show the distributions for vertically constant $K_{\rm zz}=10^8~{\rm cm^2~s^{-1}}$ for reference. We have assumed a Jupiter mass planet with a solar composition atmosphere, $T_{\rm eq}=800~{\rm K}$, and 1 Gyr age.}
}
\label{fig:Kzz_effect}
\end{figure}
While we have assumed the solar spectrum for the photochemical modeling, different stellar spectra could affect the results, especially for the photochemical depletion of NH$_3$.
Indeed, \citet{Hu21} studied photochemistry in three temperature and cold exoplanets, K2-18b, PH2 b, and Kepler-167 e, and found that photodissociation of NH$_3$ is insignificant on planets around M stars as compared to those around G/K stars (see also \citealt{Baeyens+22} for the effects of stellar spectral type in pseudo-2D photochemical simulations). 
This was attributed to the low UV flux of M-type stars at $200$--$230~{\rm nm}$ that mainly cause the photodissociation of NH$_3$ (\citealt{Hu21}, see also Section \ref{sec:result_age}).
Figure \ref{fig:stellar_effect} shows the stellar spectra of G-, K-, and M-type stars, where we use a solar spectrum of \citet{Gueymard18} as an analog of a G-type star, the spectrum of HD40307 \citep[$T_{\rm eff*}=4867~{\rm K}$, $R_{\rm *}=0.716R_{\rm \odot}$,][]{Stassun+19} as an analog of a K-type star, and GJ1214 as an analog of a M-type star \citep[$T_{\rm eff*}=3250~{\rm K}$, $R_{\rm *}=0.215R_{\rm \odot}$,][]{Cloutier+21}.
The latter two spectra are measured by the MUSCLES Treasury Survey\footnote{\url{https://archive.stsci.edu/prepds/muscles/}} \citep{France+16,Youngblood+16,Loyd+16}.
The cooler the star is, the weaker UV flux from $200$--$230~{\rm nm}$, which potentially lessens NH$_3$ depletion by photodissociation. 

To investigate the effects of stellar spectral type, we perform additional photochemical calculations for G-, K-, and M-type stars shown in the top panel of Figure \ref{fig:stellar_effect}.
To isolate the effects of stellar spectral type on photochemistry, we use the same \emph{P--T} profile of a Jupiter-mass planet with $T_{\rm eq}=600~{\rm K}$ at $1~{\rm Gyr}$ \citep[for the effects of stellar spectral type on \emph{P--T} profiles, see][]{Molliere+15,Fortney+20}.
We also adjust the orbital distance $d$ so that the stellar flux on the top of atmosphere is consistent with the assumed equilibrium temperature, as
\begin{equation}
    d=\frac{1}{2}R_{\rm *}\left( \frac{T_{\rm eff*}}{T_{\rm eq}}\right)^2.
\end{equation}

We confirm that NH$_3$ could be more stable against photodissociation at cooler stars.
The bottom panel of Figure \ref{fig:stellar_effect} shows the vertical distributions of NH$_3$ and HCN on the planet around G-, K-, and M-type stars.
As expected, NH$_3$ retains the quenched abundance until $\sim{10}^{-5}~{\rm bar}$ for K- and M-type stars, which is approximately an order of magnitude lower than the pressure level where photodissociation takes place for G-type stars.
Thus, we suggest that warm planets around K-type and M-type stars are ideal targets for detecting NH$_3$, as the photodissociation of NH$_3$ would be relatively insignificant.


\subsection{Previous Observations of NH$_3$ on Exoplanets}
Thus far there have been three exoplanets for which NH$_3$ absorption was reported: the canonical transiting hot Jupiter HD 209458b \citep[$T_{\rm eq}=1484~{\rm K}$,][]{Giacobbe+21} and warm Jupiters WASP-80b \citep[$T_{\rm eq}=817~{\rm K}$,][]{Carleo+22} and WASP-69b \citep[$T_{\rm eq}=963~{\rm K}$,][]{Guilluy+22}, the latter two planets fall in the temperature regime where NH$_3$ features are expected to be relatively large (see Section \ref{sec:observation}).
All of the above detections were accomplished by high-resolution transmission spectroscopy.
For HD209458 b, \citet{MacDonald&Madhusudhan17a} also reported the detection of NH$_3$ from  low-resolution transmission spectroscopy.
\citet{Giacobbe+21} detect NH$_3$ by cross-correlating the observations with synthetic spectra assuming a vertically constant NH$_3$ abundance of $1.3\times{10}^{-4}$, while \citet{MacDonald&Madhusudhan17a} constrained the NH$_3$ abundance to ${10}^{-8}$--$2.7\times{10}^{-6}$.
Our simulations of a Jupiter-mass planet at $1~{\rm Gyr}$ yield NH$_3$ abundances of $\sim{10}^{-6}$ at the pressure level of interest for $T_{\rm eq}=1400~{\rm K}$ (Figure \ref{fig:VULCAN_result}), which appears to be consistent with the constraint by \citet{MacDonald&Madhusudhan17a}.
However, we need a more careful analysis for planet specific comparisons, as the eddy diffusion coefficient $K_{\rm zz}$ and intrinsic temperature $T_{\rm int}$ could differ from planet to planet.
This is beyond the scope of this study.
Since the quenched NH$_3$ abundance is sensitive to the intrinsic temperature, the potential detection of NH$_3$ may provide observational tests on the suggested hot deep interiors of hot Jupiters \citep{Thorngren+19, Fortney+20}.

\begin{table*}[t]
  \caption{\rev{Summary of the impacts of planetary properties on NH$_3$ and HCN abundance and their detectability.}}\label{table:2}
  \centering
  \begin{tabular}{l  l} \hline
     Planetary parameters & Consequence\\ \hline \hline
     NH$_3$ & \\
     \hline \hline
    System age & Young (say $\la1$ Gyr) age leads to the quenched NH$_3$ abundance being lower than bulk N abundance.\\
    & The effect on feature strengths is moderate owing to compensation between NH$_3$ depletion and large scale height.\\
    Planetary mass & High mass (say $\ga1M_{\rm J}$) leads to a quenched NH$_3$ abundance lower than the bulk N abundance.\\
    & High mass results in weak NH$_3$ features in transmission spectra owing to NH$_3$ depletion and small scale height.\\
    Equilibrium temperature & The quenched NH$_3$ abundance is insensitive to the equilibrium temperature at $T_{\rm eq}<1000~{\rm K}$.\\
    & The NH$_3$ tends to be depleted by thermochemical conversion to N$_2$ at $T_{\rm eq}\ga 1000~{\rm K}$. \\
    & The NH$_3$ features in both transmission and emission spectra are relatively strong at $T_{\rm eq}\sim 400$--$1000~{\rm K}$.\\
    Atmospheric metallicity & High metallicity (say $\ga10\times$ solar) leads to the quenched NH$_3$ abundance being lower than the bulk N abundance. \\
    & The quenched NH$_3$ abundance only weakly depends on metallicity owing to the conversion of additional N to N$_2$.\\
    & High metallicity weakens NH$_3$ features in both transmission and emission spectra. \\
    Eddy diffusion & The quenched NH$_3$ abundance is insensitive to $K_{\rm zz}$. \\
     & Weak eddy diffusion causes the depletion of NH$_3$ owing to thermochemical conversion and photodissociation.\\
     Stellar spectral type & Cool stars with weak NUV flux mitigate the photodissociation of NH$_3$.\\

\hline \hline
     HCN & \\
     \hline \hline
    System age & HCN abundance in the upper atmosphere is nearly the same as the quenched abundance of NH$_3$ at deeper layers.\\
    & Young age reduces HCN abundance due to the depletion of NH$_3$.\\
    Planetary mass & High mass reduces the HCN abundance due to the depletion of NH$_3$.\\
    Equilibrium temperature & The HCN abundance follows the trend of quenched NH$_3$, except for high $T_{\rm eq}$ where CH$_4$ is significantly depleted.\\
    Atmospheric metallicity & High metallicity significantly depletes HCN owing to the depletion of CH$_4$. \\
    Eddy diffusion & Strong eddy diffusion lowers the pressure level of HCN formation. \\
     & HCN features get maximized at intermediate value ($K_{\rm zz}=10^{8}~{\rm {cm}^2~s^{-1}}$ in our case) of $K_{\rm zz}$.\\
     Stellar spectral type & Cool stars with weak NUV flux lowers the pressure level of HCN formation.\\

     \hline \hline
     Observational suggestions & \\
     \hline \hline
     Wavelength & Transmission spectra show NH$_3$ features of $\sim 50~{\rm ppm}$ at 1.5, 2, and 11 $\rm{\mu}m$ for Jupiter-mass planets around Sun-like stars.\\
     & Emission spectra show NH$_3$ features at 6 and 11 $\rm{\mu}m$ for the same conditions. The latter could be $>300~{\rm ppm}$.\\
     & HCN leaves strong features only at 14 $\rm{\mu}m$ in both transmission and emission spectra for solar composition.\\
     Temperature range & Planets with $T_{\rm eq}\sim 400$--$1000~{\rm K}$ would be suitable for detecting NH$_3$.\\
     Strategy & Emission spectrum searching for $11~{\rm {\mu}m}$ NH$_3$ feature is suitable for Jupiter-mass planets with small scale height.\\
     & Transmission spectrum might be able to find $1.5$ and $2~{\rm {\mu}m}$ features for Saturn-mass planets with large scale height.\\
     & HCN detection may indicate atmosphere significantly deviated from solar composition, such as high C/O ratio.
  \end{tabular}
\end{table*}

While we have predicted that warm exoplanets are favorable targets for searching for NH$_3$, no previous studies \rev{with low-resolution spectroscopy} have reported the detection of NH$_3$ for warm exoplanets. 
This could be due to the prevalence of clouds and hazes in warm exoplanetary atmospheres that mute the absorption features of gaseous molecules (e.g., \citealt[][]{Kreidberg+14,Kreidberg+18,Kreidberg+20,Knutson+14,Wakeford+17,Chachan+19,Chachan+20,Libby-Roberts+20,Mikal-Evans+21,Alam+22}, see \citealt{Dymont+21} for a recent compilation).
Since cloud and haze opacity tends to decrease with increasing wavelength, future observations at longer wavelengths would have better chances to detect NH$_3$, such as through the detection of $11~{\rm {\mu}m}$ feature.
Another possible cause of the nondetection of NH$_3$ could be deep atmospheres being much hotter than those predicted by thermal evolution models for non-synchronized exoplanets.
For example, \citet{Benneke+19} reported the nondetection of NH$_3$ on warm sub-Neptune GJ3470b, while the planet is known to have a non-zero eccentricity of $e\sim 0.1$ \citep{Kosiarek+19}.
Such non-zero eccentricity may yield hot deep atmospheres through tidal heating and affect upper atmospheric compositions \citep{Agundez+14b,Fortney+20}.

\rev{One of the interesting possibilities for the lack of NH$_3$ detection is that many warm exoplanets may have atmospheric compositions considerably different from solar composition. For example, if warm exoplanets typically have high metallicity atmospheres, the detection of NH$_3$ becomes more challenging, as shown in Section \ref{sec:observation}. The prevalence of high-metallicity atmospheres may be compatible with the mass-metallicity relation of warm Jupiters \citep{Thorngren+16}, planet formation models \citep[e.g.,][]{Fortney+13,Venturini+16,Cridland+20}, and several atmospheric observations of giant exoplanets \citep{Wakeford+18,Carone+21,ERS+22,ERS+22_NIRISS,ERS+22_NIRCam,ERS+22_G395,ERS+22_PRISM,ERS+22_SO2,Bean+23}. A planet may also have a subsolar N/O ratio if the solid accretion inside the N$_2$ snowline determines the atmospheric composition (see e.g., Figure 1 of \citetalias{Ohno&Fortney22a}), which would also lower the NH$_3$ abundance as compared that for solar N/O with the same atmospheric metallicity.}

\subsection{\rev{Vertical Variation of Eddy Diffusion Coefficient}}
\rev{While we have assumed vertically constant $K_{\rm zz}$ so far, in reality, the eddy diffusion coefficient likely has a vertical variation \citep[e.g.,][]{Zhang&Showman18a}.
The vertically-varying $K_{\rm zz}$ potentially affects our results at some points; for example, it may promote thermochemical conversion of NH$_3$ to N$_2$ at deep atmosphere because of low $K_{\rm zz}$, as seen in \citet{Moses+21}.
We can test the effect of vertically variable $K_{\rm zz}$ by assuming $K_{\rm zz}=K_{\rm zz,1~{\rm bar}}(P/1~{\rm bar})^{-0.5}$. Following \citet{Moses+21}, we also set $K_{\rm zz}=10^{10}~{\rm {cm}^2~s^{-1}}$ at $P>100~{\rm bar}$ and the maximum value of $K_{\rm zz}={10}^{11}~{\rm {cm}^2~s^{-1}}$.
Figure \ref{fig:Kzz_effect} shows the photochemical calculation results for various values of $K_{\rm zz,1~{\rm bar}}$. 
The vertical variable $K_{\rm zz}$ tends to facilitate the thermochemical conversion of NH$_3$ to N$_2$ owing to the weak vertical mixing in the deep atmosphere, while it mitigates the NH$_3$ photodissociation thanks to the strong mixing in the upper atmosphere.
The HCN abundance in the upper atmosphere reflects the quenched NH$_3$ abundance as in the constant $K_{\rm zz}$ case, while HCN abundance at lower atmosphere could be higher  compared to constant $K_{\rm zz}$ cases because downward diffusive transport becomes slower as the pressure increases.
The vertical variation of $K_{\rm zz}$ and its strength have remained poorly constrained by observations \citep{Kawashima&Min21}.
Future observations of disequilibrium species more sensitive to the quench point than NH$_3$ is, such as CO, would help to probe $K_{\rm zz}$ in deep atmospheres \citep[e.g.,][]{Miles+20,Mukherjee+22b}.
}

\subsection{\rev{Effects of Day-Night Temperature Contrast}}
Our study has relied on a 1D framework, whereas real exoplanetary atmospheres are in 3D and \rev{may show strong temperature contrasts between daysides and nightsides \citep[e.g.,][]{Perez-becker&Showman13,Komacek&Showman16}}; however, we do not anticipate that this is a major drawback in this case.
Several studies have investigated how horizontal temperature variations and horizontal winds affect the chemistry of exoplanetary atmospheres.  These include 3D GCMs \citep{Cooper&Showman06,Mendonca+18,Drummond+18,Drummond+18b,Drummond+20}, equatorial 2D models \citep{Tsai+21_2D}, and pseudo-2D models that rotate a 1D model along the equator \citep{Agundez+12,Agundez+14,Moses+21,Baeyens+21,Baeyens+22}.
According to the pseudo-2D works by \citet{Moses+21} and \citet{Baeyens+22}, the NH$_3$ abundance is almost invariant with longitudes at warm exoplanets with $T_{\rm eq}\la1000~{\rm K}$, except for the pressure level of $\la {10}^{-3}$--${10}^{-4}~{\rm bar}$ where the photodissociation takes place. 
\rev{
\citet{Drummond+20} performed 3D global circulation simulations on hot Jupiters HD 189733b and HD 209458b using a 3D GCM that implements a full kinetic chemistry based on a reduced chemical network of \citet{Venot+19} and showed that NH$_3$ abundance is vertically and longitudinally uniform in both planets, though they did not include photochemistry.
}
\rev{These longitudinal independence} could be attributed to the long thermochemical conversion timescale of NH$_3$, $>{10}^{15}~{\rm s}$ at $1000~{\rm K}$ \citep[see Figure 4 of][]{Tsai+18}, that is much longer than typical horizontal transport timescales.
In the deep atmosphere where the chemical quenching takes place, on the other hand, the atmospheric \emph{P--T} profile is invariant with longitude because of the long radiative timescale.
Thus, we anticipate that the NH$_3$ spatial distribution in real atmospheres is reasonably approximated by the 1D framework, especially for warm exoplanets that are optimum for detecting NH$_3$.

\rev{
\subsection{Observability of N$_2$ molecule}
In \citetalias{Ohno&Fortney22a} and this study, we have focused on NH$_3$ and HCN since N$_2$, the remaining main N reservoir, in general has negligibly low opacity in visible and infrared wavelengths.
However, N$_2$ actually has moderate opacity at extremely hot temperature, say $>4000~{\rm K}$.
High resolution spectroscopy might still have a chance to detect N$_2$ if it abundantly exists in hot thermosphere.
In the context of terrestrial exoplanets with N$_2$ dominated atmospheres, \citet{Schwieterman+15} suggested that N$_2$ can be detected from the absorption feature of $\rm(N_{\rm 2})_{\rm 2}$ dimer produced by N$_2$--N$_2$ collisions, though it unlikely affects the observable spectra of giant planets with much lower N$_2$ abundances.
N$_2$--H$_2$ collision-induced absorption (CIA) opacity might still have some impacts; however, quantitative assessment is difficult as of yet owing to the limited valid range of wavelength and temperature for the available CIA absorption coefficient \citep{Karman+19}. 
}



\section{Summary}\label{sec:summary}

\rev{In this study, we have performed} a series of photochemical calculations for various values of planetary mass, age, equilibrium temperature, eddy diffusion coefficient, and atmospheric composition to explore the relation between observable NH$_3$ abundance and bulk nitrogen abundance.
\rev{Based on the photochemical calculations and the semi-analytic model of NH$_3$ quenching developed by \citetalias{Ohno&Fortney22a} (Equation \ref{eq:NH3_analytic}), we have comprehensively revealed what planetary properties act to deplete observable NH$_3$ as compared to bulk nitrogen. Table \ref{table:2} summarizes how observable abundance of NH$_3$ and HCN depend on various planetary properties and its observational implications.}
Our key findings are summarized as follows:

\begin{enumerate}


\item \rev{As shown in \citetalias{Ohno&Fortney22a}}, the vertically quenched NH$_3$ abundance is nearly identical to the bulk nitrogen abundance \emph{only} when a planet has a sub-Jupiter mass ($\la1~{\rm M_{\rm J}}$) and old age ($\ga 1~{\rm Gyr}$) under the assumption of a solar composition atmosphere. 

\item \rev{High metallicity atmospheres lead to the quenched NH$_3$ abundance being much lower than the bulk nitrogen abundance even at  sub-Jupiter planet masses} and old ages. This highlights the importance of constraining overall atmospheric metallicity for inferring bulk nitrogen abundances.

\item The semi-analytical model of \citetalias{Ohno&Fortney22a} (Equation \ref{eq:NH3_analytic}) reproduces the vertically quenched NH$_3$ abundance computed by a photochemical kinetic model (Section \ref{sec:N_map}).
Thus, our semi-analytical model would mitigate the discrepancy between the quenched NH$_3$ and bulk nitrogen abundances when inferring the bulk nitrogen abundance from an observed NH$_3$. 

\item NH$_3$ is vulnerable to photodissociation and tends to be depleted at ${10}^{-3}$--${10}^{-4}~{\rm bar}$ in clear atmospheres, depending on the equilibrium temperature and eddy diffusion coefficient.
The relatively deep pressure level of NH$_3$ photodissociation is caused by UV photons at $200$--$230~{\rm nm}$ that can penetrate to higher pressures owing to the lack of absorption opacity of other molecules at those wavelengths.


\item The photodissociation of NH$_3$ is mitigated when planets orbit around cool K and M stars because of the decreased level of NUV photons. Thus, warm planets around K and M stars would be ideal targets for searching for NH$_3$.

\item We have examined the NH$_3$ and HCN signatures in transmission spectra of warm gas giants.
For 1 Gyr Jupiter-mass planets with clear atmospheres around Sun-like stars, we have predicted that NH$_3$ would leave characteristic features of $>50~{\rm nm}$ at $1.5$, $2.1$, and $11~{\rm {\mu}m}$ in transmission spectrum for the equilibrium temperature of $T_{\rm eq}\sim 400$--$1000~{\rm K}$.

\item \rev{The emission spectra shows  NH$_3$ features at $\sim 6$ and $11~{\rm {\mu}m}$. For Jupiter-mass planets around Sun-like stars, we have predicted that the $11~{\rm {\mu}m}$ feature could be $>300~{\rm ppm}$, much larger than that expected for transmission spectra.}

\item The $11~{\rm {\mu}m}$ feature is particularly strong and would be the best signature to identify NH$_3$, as clouds and/or hazes tend to have less opacity at such long wavelengths.

\item \rev{Our analysis has suggested that it may be difficult to detect HCN by low-resolution spectroscopy if a planet has a solar composition atmosphere. This is because HCN is formed by photochemistry that takes place in the upper atmosphere far away from the pressure level typically probed by observations. The detection of HCN potentially indicate that atmospheric composition is considerably different from solar composition, such as a high C/O ratio of C/O$\gg$1.}

\end{enumerate}

\rev{
Our studies have highlighted that it is not \rerev{trivial} to robustly constrain the atmospheric nitrogen abundances of exoplanets.
The observable NH$_3$ abundance underestimates the bulk nitrogen abundance in most cases.
Thus, the observable NH$_3$ abundance is usually a lower limit of bulk nitrogen abundance, and chemical models are necessary to constrain the bulk nitrogen abundance.
This task will be aided by photochemical kinetic models \citep[e.g.,][]{Line+11,Moses+11,Venot+13,Kawashima&Ikoma18,Molaverdikhani+19,Tsai+21,Hu21} and/or retrieval models including disequilibrium chemistry effects \citep{Kawashima&Min21,Al-Refaie22}. Our semi-analytic model (Equation \ref{eq:NH3_analytic}) can provide a first estimate on the bulk nitrogen abundance from a retrieved NH$_3$ abundance.
}

\rev{
Although the robust constraint of nitrogen abundance is not easy, we expect that a lower-limit nitrogen abundance still provides some insights on planet formation.
For example, for a superstellar N/O ratio, a lower-limit N/O ratio may be able to give a lower-limit orbital distance of planet formation location because gas-phase N/O monotonically increases with the distance \citep{Piso+16}.
Alternatively, if planetesimals and/or pebble accretion mainly determine the atmospheric composition, which could be inferred from refractory element abundances in the atmosphere, such as sulfur and alkali metals \citep[e.g.,][]{Schneider&Bitsch21b,Turrini+22,Hands&Helled22,Pacetti+22}, an N/O ratio higher than $\sim0.1\times$ stellar value may indicate the planet formation at extremely cold region where N$_2$ ice could be frozen, as suggested for Jupiter in our Solar System \citep{Owen+99,Oberg&Wordsworth19,Bosman+19,Ohno&Ueda21}.
Given the unprecedented precision and wavelength coverage of JWST \citep{ERS+22,ERS+22_NIRISS,ERS+22_NIRCam,ERS+22_G395,ERS+22_PRISM}, it will be important to continue to assess what we can learn about planet formation from various elemental ratios of exoplanetary atmospheres.
}

\section*{Acknowledgements}
We are grateful to anonymous reviewer for insightful comments that greatly improved the quality of this paper.
We thank Shang-Min Tsai for helpful instructions on the analysis of VULCAN.
We also thank Neel Patel, Masahiro Ikoma, Xinting Yu, Ben Lew, Eliza Kempton, Yuichi Ito, Yui Kawashima, Shota Notsu, Tatsuya Yoshida, and Akifumi Nakayama for fruitful discussions.
This work benefited from the 2022 Exoplanet Summer Program in the Other Worlds Laboratory (OWL) at the University of California, Santa Cruz, a program funded by the Heising-Simons Foundation.
Most of numerical computations were carried out on PC cluster at Center for Computational Astrophysics, National Astronomical Observatory of Japan.
K.O. was supported by JSPS Overseas Research Fellowship.  J.J.F. is supported by
an award from the Simons Foundation.

\bibliography{references}
\end{document}